\documentclass[final,5p,times,twocolumn]{elsarticle}

\usepackage{amssymb}
\usepackage{amsmath}
\usepackage{siunitx}
\usepackage{upgreek}
\usepackage{booktabs} 

\journal{Acta Materialia}

\begin{document}

\begin{frontmatter}



\title{Spatially resolved in-situ characterisation of competing martensitic transformation pathways during nanoscratch in 316H Stainless Steel}

\author[a]{A. Kareer\corref{cor1}}
\ead{anna.kareer@materials.ox.ac.uk}
\cortext[cor1]{Corresponding author}
\author[b]{R.W. Kerr}
\author[c]{D. Craven} 
\author[d]{A.V. Davydok}
\author[d]{C. Krywka}
\author[e]{D.M. Collins\corref{cor2}}
\ead{dmc51@cam.ac.uk}
\cortext[cor2]{Corresponding author}

\affiliation[a]{organization={Department of Materials, University of Oxford, Parks Road, Oxford, OX1 3PH, United Kingdom}}
\affiliation[b]{organization={United Kingdom Atomic Energy Authority, Culham Campus, Abingdon OX14 3DB,
United Kingdom}}
\affiliation[c]{organization={Department of Materials, Imperial College London, Prince Consort Road, London, SW7 2BP, United Kingdom}}
\affiliation[e]{organization={Department of Materials Science \& Metallurgy, University of Cambridge, 27 Charles Babbage Road, Cambridge, CB3 0FS, United Kingdom}}
\affiliation[d]{Institute of Materials Physics, Helmholtz-Zentrum Hereon GmbH, Outstation at DESY, Notkestr. 85, Hamburg, 22607, Germany}

\begin{abstract}

Localised surface deformation beneath frictional contacts generates a tribolayer whose microstructure and properties differ from the bulk. In austenitic stainless steels, this tribolayer forms through two competing martensitic transformation pathways. Here, these pathways are isolated in 316H stainless steel using in-situ synchrotron X-ray nanodiffractometry combined with nanoscratch testing, which together yield spatial maps of the evolving strain field beneath a single sliding asperity. Finite element modelling interprets the resulting distribution of martensitic phases, revealing a pressure driven pathway selection where hydrostatic compression ahead of the contact suppresses $\alpha'$  formation and favours the $\gamma \rightarrow \varepsilon$  transformation, while lateral sliding relieves this constraint and introduces a shear strain driving $\varepsilon \rightarrow \alpha'$ , producing an overall sequential $\gamma \rightarrow \varepsilon \rightarrow \alpha'$ pathway in the tribolayer. Where hydrostatic constraint persists, $\varepsilon$-martensite is retained; where material piles up and is unconstrained above the surface, the transformation proceeds directly to $\alpha'$. The $\gamma$-austenite adjacent to $\alpha'$- martensite shows elevated dislocation density, indicating that $\alpha'$ formation is accommodated by plastic deformation in the surrounding matrix. This distinction could explain differences in galling performance among iron-based and cobalt-based hardfacing alloys, where the $\varepsilon$-martensite forming cobalt alloys offer superior galling resistance. The methodology presented resolves transient microstructural states inaccessible to static measurements of macroscale, multiple asperity contacts, establishing a route to mechanistic insight across tribological phenomena more broadly.

\end{abstract}





\end{frontmatter}
\section{Introduction}
\label{sec1}

Under dry sliding conditions, metallic materials undergo severe plastic deformation resulting in the formation of localised surface layer, whose structure and properties differ markedly from the bulk material \cite{rigney1997comments, Rigney1984}.  This layer, termed the \textit{tribolayer}, evolves progressively under continued surface deformation during wear, through mechanisms including dislocation rearrangement, grain refinement, dynamic recrystallisation, oxidation, and strain induced phase transformation \cite{greiner2018origin, greiner2016sequence, emge2009effects, yao2012correlation, hubner2003phase, rau2021high}. The resulting structure governs the macroscopic friction and wear response of the material. Understanding the mechanisms that drive tribolayer formation and evolution is fundamental to predicting and improving the tribological performance of materials.

The tribolayer is particularly relevant to galling; a severe form of adhesive wear in which metal surfaces weld together under high contact pressures and slow sliding regimes \cite{hutchings2017tribology}. It initiates via the formation of localised adhesive junctions between microscale surface asperities, and progresses via junction growth under plastic deformation, eventually forming a strong, cold weld. As sliding continues, subsurface plastic strain accumulates and localises in the vicinity of these junctions, resulting in severely roughened surfaces, the transfer or displacement of large fragments of material and seizure \cite{astm_g40_15, hutchings2017tribology}. The severity of galling and its progression is governed by the evolving properties within the tribolayer, rather than those of the bulk material.

Nuclear valves are subjected to thermal cycling and high pressures (often $\sim$100 MPa at temperatures around \qty{300}{\degree C}), making them susceptible to seizure from galling \cite{daure2025significant}. Reliable valve operation is vital for controlling coolant systems and isolating radioactive materials in the primary circuit, hence their failure represents a critical risk, compromising the plant safety, regulatory compliance, and economic operation \cite{vikstroem1994galling}. Cobalt-based hardfacing alloys, such as the commercial alloy Stellite, are widely adopted for these applications owing to their superior wear and galling resistance; however, cobalt containing wear debris travel into the reactor core and become activated under the neutron flux \cite{inglis1992performance, bowden2019understanding}. Under the ALARA (as low as reasonably achievable) principle, the resulting high activation levels of cobalt-60 have motivated sustained efforts to identify wear and galling resistant, cobalt-free replacement alloys \cite{vannerem88chemistry, cockeram1999development, cachon1996tribological}.  

The superior galling resistance of cobalt-based hardfacing alloys has been attributed to several competing or coexisting mechanisms, including the intrinsically low stacking fault energy of the cobalt matrix, solid solution strengthening from chromium and tungsten, and the presence of hard carbide phases that resist asperity deformation \cite{ahmed2017friction, PaulCrook1990, antony1983wear}. Under sliding contact, a distinct tribolayer has been observed in cobalt-based hardfacing alloys, in which the $\gamma$-Co matrix undergoes a deformation induced $\gamma \rightarrow \varepsilon$ martensitic phase transformation \cite{zhao2018comparative}. Among the proposed mechanisms, this martensitic transformation has been suggested as a contributing factor to galling resistance, whereby the hardened structure promotes easy shear of junctions, suppressing their growth and the accumulation of subsurface plasticity \cite{persson2003influence, persson2003antigalling}. 

Identifying an analogous mechanism in cobalt-free candidate systems is central to the design of replacement hardfacing alloys. Iron-based austenitic stainless steels represent the most widely studied matrix system for cobalt-free replacements, owing to their chemical compatibility with reactor components and their well-documented tendency for deformation induced martensitic transformation during sliding \cite{bastola2024experimental}. Despite both classes of hardfacings exhibiting a hard martensitic tribolayer, the iron-based hardfacings consistently deliver inferior galling resistance relative to cobalt-based alloys \cite{persson2003effect, bowden2019understanding, rogers2020interaction}. 

A key distinction is that while cobalt-based alloys undergo a direct $\gamma \rightarrow \varepsilon$ (fcc $\rightarrow$ hcp) martensitic transformation, the strain induced transformation in austenitic stainless steels results in the formation of a bct structured $\alpha'$-martensite \cite{Shen2019Carbon} phase. The latter can proceed via a direct $\gamma \rightarrow \alpha'$ (fcc $\rightarrow$ bct) pathway, under conditions of high plastic shear strain \cite{mangonon1970martensite}, or via a sequential $\gamma \rightarrow \varepsilon \rightarrow \alpha'$ (fcc $\rightarrow$ hcp $\rightarrow$ bct) pathway, whereby an intermediate $\varepsilon$-martensitic is formed via the accumulation of stacking faults on close-packed \{111\} planes \cite{olson1972mechanism}. The relative activity and stability of each transformation pathway is governed by the stacking fault energy of the alloy  and the imposed deformation conditions \cite{perdahciouglu2012macroscopic}. Critically, the resulting martensitic phases carry distinct volumetric character based on their crystallographic structure, the $\gamma \rightarrow \varepsilon$ transformation (fcc $\rightarrow$ hcp) is approximately volume conserving, whereas both the direct and sequential routes that form $\alpha'$ (bct) introduce a positive dilatation relative to the parent austenite (fcc) \cite{greenwood1965deformation}.  
The crystallographic differences between the deformation induced transformations in cobalt- and iron-based hardfacings have not been explicitly linked to the near-surface deformation generated during sliding, nor have they been established as a mechanistic explanation for the performance gap between these alloy classes. Resolving this distinction using conventional wear testing approaches is challenging due to the dynamic nature of the transformation and the microstructural evolution during sliding. post-mortem characterisation of wear surfaces in austenitic stainless steels and commercial iron-based hardfacing alloys, only show the resultant $\alpha'$-martensite \cite{kim2000temperature, rogers2024mechanisms, rogers2020interaction} and while both $\alpha'$-martensite and $\varepsilon$-martensite have been resolved within spatially distinct, subsurface regions using TEM \cite{carrington2024evolution}, the transformation pathway can only be inferred from these static measurements. In-situ X-ray diffraction measurements during macroscale wear testing of stainless steel 316 show that $\varepsilon$-martensite forms at the early stages of wear testing at distinct locations, and is converted to $\alpha'$-martensite, directly following the sequential $\gamma \rightarrow \varepsilon \rightarrow \alpha'$ transformation pathway \cite{emurlaev2022friction}. However, the inherent inhomogeneity of the multi-asperity contact during macroscale wear means that the mechanisms driving pathway selection remain unknown and cannot be directly correlated to the local deformation; such mechanistic understanding requires the use of single asperity experiments as a model system for wear \cite{Bhushan1996Contact, Cappella2022Editorial:, Stoyanov2017Scaling}. 

Recent developments in in-situ nanomechanical testing methods overcome these limitations, by resolving the temporal evolution of mechanical contacts, at length scales relevant to single asperity interactions \cite{jacobs2019insights}. In-situ synchrotron X-ray diffraction measurements of thin film nanoindentation has resolved the evolving stress distributions and microstructural changes beneath the indentation, attributes that are inaccessible to post-mortem characterisation \cite{zeilinger2016situ}. Nanoscratch testing offers a platform with sufficient control and resolution to reproduce the deformation fields from a single sliding asperity, with well-defined contact mechanics \cite{pethica2023nanoindentation,kareer2016existence}. Combining nanoscratch with post-mortem cross-sectional electron backscatter diffraction, the localised deformation fields within the tribolayer have been spatially resolved and directly linked to the dynamic contact mechanics during sliding \cite{kareer2020scratching, kareer2025localised}

In this work, in-situ synchrotron X-ray diffraction is combined with nanoscratch testing to resolve the martensitic transformation pathways within the tribolayer of austenitic stainless steel 316H, both spatially and dynamically during the test. 316H is studied as a model matrix system for iron-based hardfacing alloy development. Spatially resolved diffraction maps are acquired at successive stages of scratch progression and are interpreted alongside finite element simulations of the contact stress and plastic strain fields beneath a well-defined, single asperity contact. This enables the transformation pathway, the evolving near-surface deformation, and the corresponding stress distribution to be directly correlated throughout tribolayer formation and evolution.

\section{Methods}
\label{sec2}
\subsection{In-situ nanoscratch X-ray diffraction}

Samples of of 316H stainless steel were prepared by mechanically grinding a chamfered edge on a \qty{200}{\milli\meter} $\times$ \qty{50}{\milli\meter} $\times$\ \qty{15}{\milli\meter} bar, leaving a polished edge with thickness of \qty{200}{\micro\meter}. A plasma Focussed Ion Beam (Thermo-Fisher Helios G4-CXe PFIB) was used to prepare a \qty{300}{\micro\meter} $\times$ \qty{200}{\micro\meter} lamella with \qty{50}{\micro\meter} thickness (parallel to the beam direction). 

\begin{figure}
    \centering
    \includegraphics[width=1\linewidth]{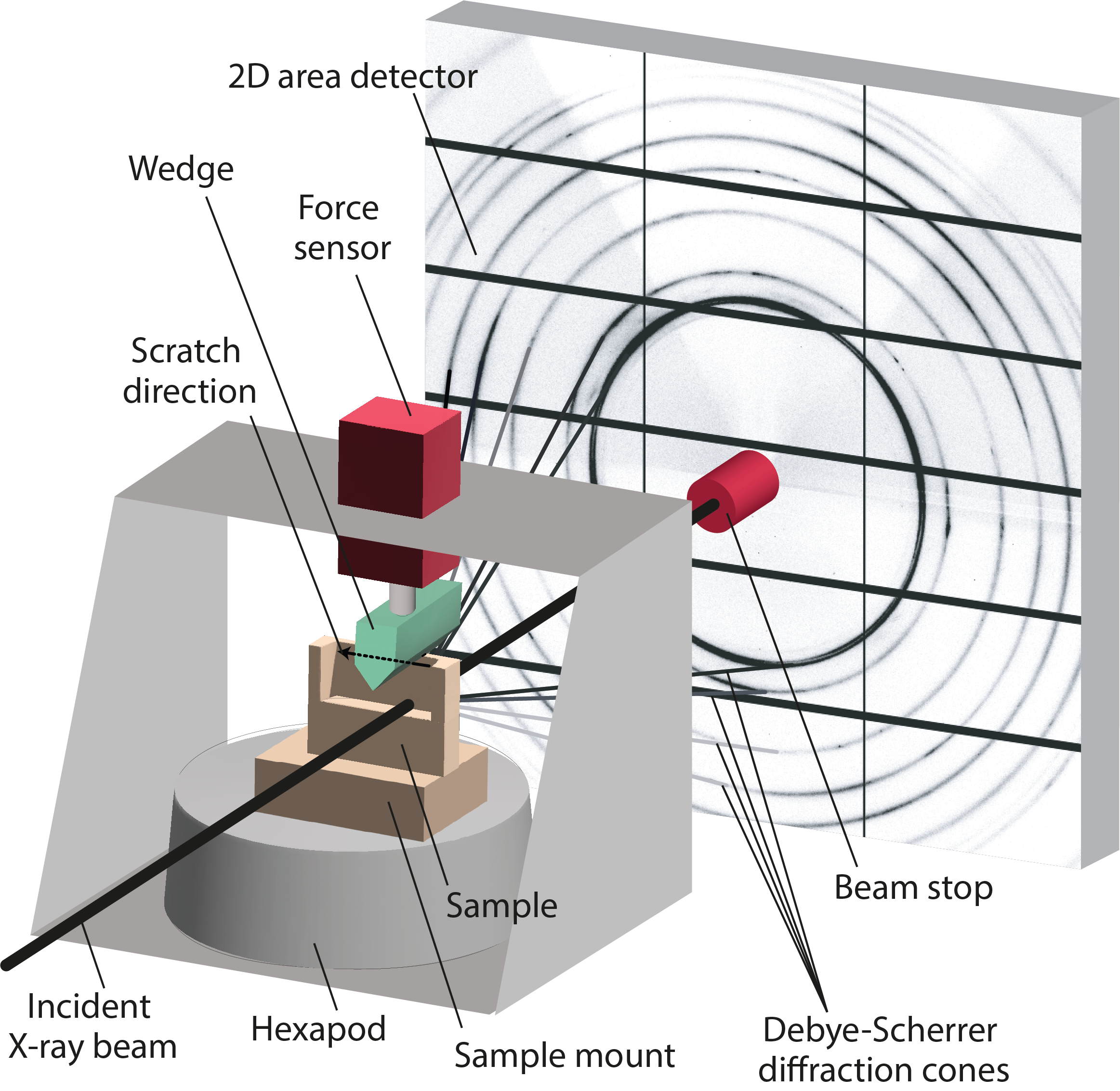}
    \caption{Schematic illustration of the in-situ X-ray diffraction on the Nanofocus Endstation, P03, at PETRA III, DESY, equipped with an indentation system, employed here to before scratch testing.   }
    \label{fig:setup}
\end{figure}

The transmission X-ray nanodiffraction experiment was performed on the Nanofocus Endstation of the P03 MiNaXS beamline at the synchrotron light source, PETRA III at DESY (Hamburg, Germany). 
As shown in Fig. \ref{fig:setup}, the beamline was equipped with a custom built indentation system. The setup, originally developed for static nanoindentation, was modified in this experiment to perform in-situ nanoscratch testing \cite{zeilinger2016situ,TODT2020425}.
The lamellae-containing sample was mounted in a holder that was fixed onto a hexapod stage. The latter permits multi-axis alignment (translation and rotation), as well as step-wise displacement during X-ray scanning measurements. Critically, it also enabled lateral movement of the sample relative to the indenter tip, necessary to conduct the nanoscratch experiment. The system comprised a wedge-shaped diamond tip with a tip radius of $<$\qty{20}{\micro\meter}, a wedge length of $\sim$\qty{200}{\micro\meter} and $\sim$\qty{70}{\degree} opening angle. Attached to the tip was a force transducer with a digital readout. 
A monochromatic X-ray beam at \qty{19.85}{\kilo\electronvolt} (\qty{0.6246}{\angstrom}), calibrated with a NIST LaB$_6$ powder standard, was used.
Using Kirkpatrick–Baez mirrors, the X-ray beam was focused to ($1\times 1$\,\unit{\micro\meter\squared}).
Debye-Scherrer diffraction patterns were acquired with a time acquisition of 1 second, using a DECTRIS EIGER 9M area detector, comprising a $75\times75$\,\unit{\micro\meter\squared} pixel size onto an active area of $233\times245$\,\unit{\milli\meter\squared}, at a sample-to-detector distance of $\sim185.5$\unit{\milli\meter}.
Two-dimensional diffraction maps were recorded beneath the indenter tip at intervals during the nanoscratch. The first maps were recorded on the pristine material before the indenter had made contact with the sample. The indentation step involved vertically raising the sample towards the indenter until contact was made, indicated by a fluctuation in the force reading. The sample was further raised until an indentation was made under a force of \qty{250}{\micro\newton}, where the second map was recorded. The third and fourth maps were recorded during nanoscratch intervals. The sample was moved laterally beneath the indenter to generate a scratch in steps of \qty{2}{\micro\meter}; the force was recorded for each step. Maps were recorded at a scratch length of \qty{20}{\micro\meter} and \qty{70}{\micro\meter}. The final map was acquired in the unloaded state, when the tip was removed from the sample by lowering the stage. Each map measured $100\times 60$\,\unit{\micro\meter\squared} with a step size of $1$\,\unit{\micro\meter}. The selection of a \qty{60}{\micro\meter} depth below the surface was informed by finite element simulations to ensure the sampled volume contained the deformation features of interest. As the maximum displacement range of the stage for mapping was \qty{100}{\micro\meter}, some intervals required two maps to be recorded and stitched. 

\subsection{Diffraction analysis}
\label{sec3}

Debye-Scherrer diffraction patterns were processed using DAWN \cite{Basham2015,Filik2017}. Azimuthal cake integration was performed over a \ang{30} sector to obtain 12, one-dimensional intensity profiles as a function of scattering vector $q$ for each mapped location. Peak fitting was performed using an in-house fitting method, written in MATLAB with a Gaussian function and a linear background of the form
\begin{equation} 
I(q) = A \exp\left[-\left(\frac{q-q_0}{w}\right)^2\right] + d_0 q + d_1,
\end{equation}
where $A$ is the peak amplitude, $q_0$ is the peak centre, $w$ controls the peak width, and $d_0$ and $d_1$ are coefficients of a linear background. For the partially overlapping reflection at $q$ $\approx$ 3\,\AA$^{-1}$ (see Figure \ref{fig:peaks}.b), a double-Gaussian model was employed to separate the intensity contributions of each. For every fitted reflection, a fitting window was defined, within which the peak was detected (using the signal processing toolbox in Matlab) and a non-linear least-squares fit was initialised. 

The full width at half maximum for a given reflection (FWHM$_{hkl}$) was calculated from the fitted Gaussian parameter as
\begin{equation} 
\mathrm{FWHM}_{hkl} = 2\sqrt{\ln 2}\, w,
\end{equation}
and the integrated intensity, ${\cal{I}}$ , was determined analytically as
\begin{equation} 
{\cal{I}} = A w \sqrt{\pi}.
\end{equation}

The component of the $\gamma$-phase lattice strain aligned parallel to the scratch direction, $\varepsilon_{hkl,xx}$, was determined from the diffraction peak positions of horizontal sectors corresponding to azimuthal angles \ang{0} and \ang{180}, which were averaged to obtain a single peak position per reflection. The peak positions of three \{111\}, \{200\}, and \{220\} reflections were converted from $q$ to $d$ and the lattice strain was calculated as
\begin{equation}
    \varepsilon_{hkl,xx} = \frac{d_{hkl} - d_{0,hkl}}{d_{0,hkl}}
    \label{eq:lattice_strain}
\end{equation}
where $d_{0,hkl}$ is the reference $d$-spacing for reflection \{$hkl$\}, taken as the mean $d_{hkl}$ over all pixels in the pristine (undeformed) map. Reflection-specific stresses were then obtained from
\begin{equation}
\sigma_{hkl} = M_{hkl}\,\varepsilon_{hkl,xx}
\end{equation}
where $M_{hkl}$ is the diffraction elastic constant for the corresponding reflection calculated from the Kröner model \cite{kroner1958berechnung}. The stresses from the three reflections were combined by averaging, and the resulting mean stress was converted to an effective $\gamma$-austenite strain, $\varepsilon_{\text{bulk}}$, using
\begin{equation}
\varepsilon_{\text{bulk}} = \bar{\sigma}/E.
\end{equation}
where $E = 193$\,GPa for 316 stainless steel \cite{banerjee2023finite}. Values of $\varepsilon_{\text{bulk}}$ are primarily studied here in the $x$ direction; these are herein given as $\varepsilon_{xx}$, through this manuscript.

The total dislocation density was determined from diffraction peak broadening using a modified Williamson–Hall approach. For each pixel, the FWHM of the $\gamma$-austenite \{111\}, \{200\}, and \{220\} reflections was obtained from peak fitting and corrected for instrumental broadening using a Stokes-type correction. The modified Williamson–Hall relation,
\begin{equation}
\label{eq:williamson hall}
\frac{\beta \cos \theta}{d} = \frac{K\lambda}{D} + m \sqrt{C_{hkl}}
\end{equation}
was applied, where $\beta$ is the instrumentally corrected $FWHM$, $\theta$ is the Bragg angle, $d$ is the $d$-spacing of reflection \{$hkl$\}, $K$ is the Scherrer shape constant (taken as 0.9), $\lambda$ is the X-ray wavelength, $D$ is the mean crystallite size, $m$ is the slope of a per-pixel weighted least squares regression of the left-hand side of Eq. \ref{eq:williamson hall} against $\sqrt{C_{hkl}}$ across the three reflections, and $C_{hkl}$ is the dislocation contrast factor \cite{ungar1999contrast}, calculated using the single crystal elastic constants, $C_{11}$, $C_{12}$ $C_{44}$ \cite{agius2020microstructure}. 

The dislocation density, $\rho$, was calculated using
\begin{equation}
\rho = \frac{4 m^2}{\pi M b^2}
\end{equation}
where $M$ is a dimensionless constant related to the effective radius of the dislocation strain fields (taken as 2), and $b$ is the magnitude of the Burgers vector, equal to $a_0/\sqrt{2}$, with $a_0 = 3.596\,\AA{}$ the $\gamma$-austenite lattice parameter.

Confidence intervals for the fitted parameters were obtained from the non-linear regression. Uncertainties in the calculated metrics were obtained by propagation of the fitted parameter uncertainties using the delta method. For completeness, all uncertainties are reported in Appendix \ref{appendix 1} and correspond to one standard deviation $(1\sigma)$.  

\subsection{Finite element model}
\label{FE model method}

Finite element simulations were performed using Abaqus 2022 (Standard).  A simplified two-dimensional plane stress model of the lamella was constructed using quadratic reduced-integration elements (CPS8R). Large deformation kinematics were enabled (NLGEOM = ON). The indenter was modelled as an analytical rigid wedge matching the experimental geometry, with an opening angle of 70$^\circ$. Surface-to-surface contact was defined using a hard normal contact formulation, with a Coulomb friction coefficient of 0.15 \cite{frictioncoeffgangopadhyay1993friction}. The simulation consisted of indentation to a maximum depth of \qty{2}{\micro\meter}, followed by lateral translation of \qty{70}{\micro\meter} at constant depth to simulate scratching, followed by  subsequent unloading. The sample measured $\qty{300} \times \qty{150}{\micro\meter}^{2}$; the bottom, left and right edges of the sample were fully constrained. The 316H stainless steel was modelled as isotropic elastic–plastic. Elastic properties comprised a Young’s modulus of 193\,GPa and Poisson’s ratio of 0.3 \cite{banerjee2023finite}. Plastic behaviour was defined using a tabulated true stress–plastic strain curve (see Table \ref{tab:plasticity} in Appendix \ref{appendix 2}). The isotropic hardening formulation was adopted since the aim was to resolve the subsurface strain fields generated from the single asperity contact, rather than simulate pile-up evolution that would require a kinematic hardening model. A biased mesh was employed, with element sizes refined to \qty{0.3}{\micro\meter} in the contact region and gradually coarsened to \qty{10}{\micro\meter} away from the surface. All dimensions were defined in \qty{}{\micro\meter} and stresses in MPa.

\section{Results}
\label{sec5}

\subsection{Nanoscratch experiment}
\label{nanoscratch experiment}
\begin{figure}[h]
    \centering
    \includegraphics[width = 1.00\linewidth]{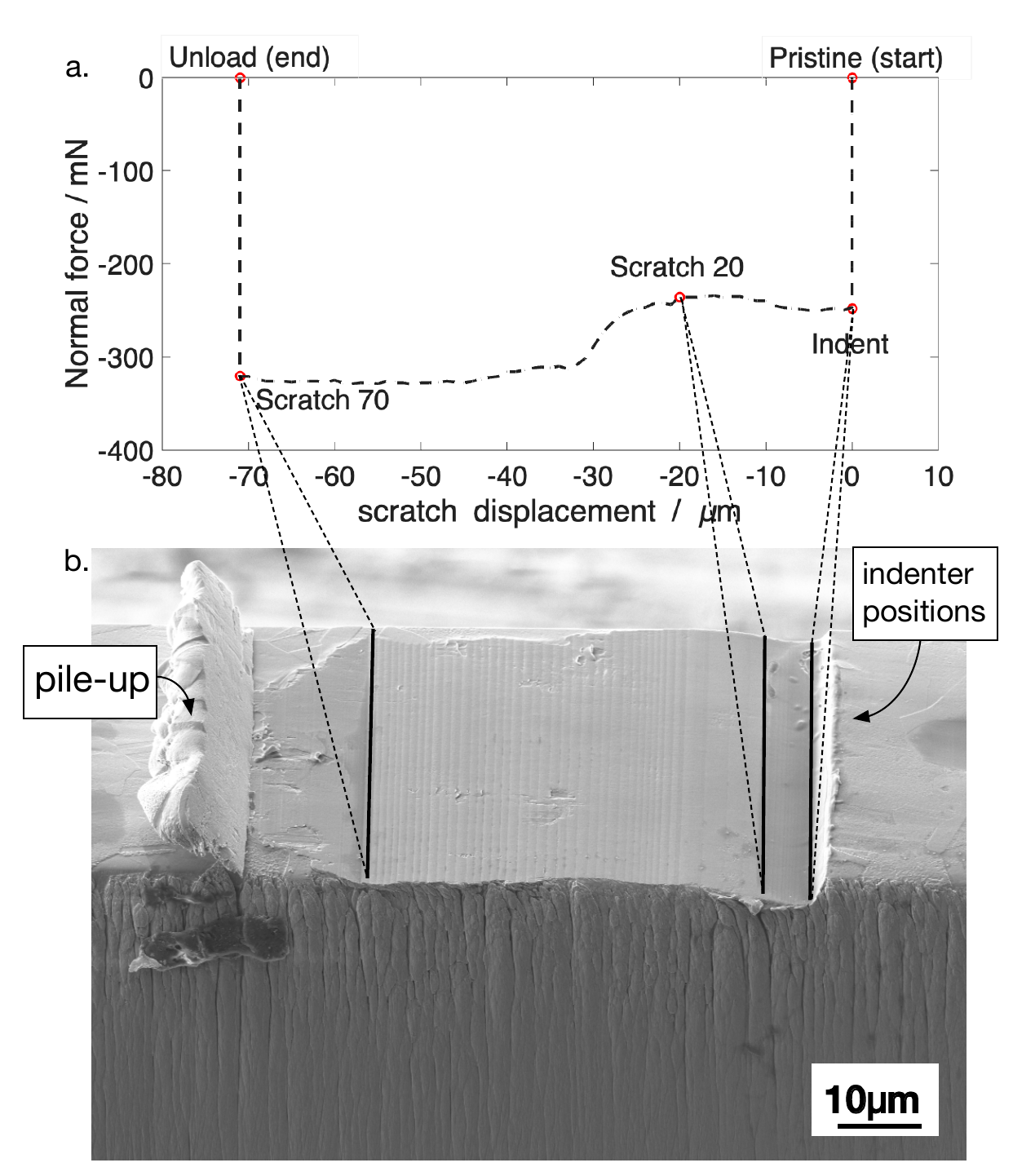}
    \caption{Nanoscratch experiment overview. (a) Normal force as a function of scratch displacement recorded during the nanoscratch test. Red circles indicate the increments at which two-dimensional diffraction maps were collected (labelled Pristine, Indent, Scratch 20, Scratch 70, and Unload). Dashed lines connect each increment to its corresponding position in (b). (b) SEM micrograph of the scratch track after the experiment. Solid lines denote the approximate position of the wedge indenter tip at each measurement increment. Piled-up material has detached from the surface at the end of the scratch track.}
    \label{fig:experiment}
\end{figure}

The deformation history and the evolving contact mechanics throughout the in-situ nanoscratch experiment is shown in Figure \ref{fig:experiment}. The normal load is plotted as a function of the scratch displacement in Figure \ref{fig:experiment}a, showing the in-situ intervals at which the diffraction maps were collected. The SEM micrograph (Figure \ref{fig:experiment}b) shows the residual surface morphology including a side-on view of the lamella from which the scratch penetration depth and pile-up can be observed. At the surface, the corresponding instantaneous position of the wedge indenter for each interval is given by the black solid lines. As neither the normal nor lateral displacement of the tip was directly measured, the scratch displacement ($x$-axis of Figure \ref{fig:experiment}a) represents the imposed stepwise lateral stage displacements. The SEM micrograph can be used to infer the normal displacement for each interval. While the experiment was nominally conducted in displacement control, the observed variable penetration depth along the scratch track, and the discrepancy between the indenter positions in Figure \ref{fig:experiment}a and \ref{fig:experiment}b, indicate the high levels of compliance in the indentation system. The load-displacement data can be described by evolving contact mechanics; an initial reduction in normal force, between the indentation and \qty{20}{\micro\meter} scratch interval, corresponds to the simultaneous loss of contact area and increase in lateral resistance during the transition from indentation to scratching \cite{brazil2021contribution}. To sustain scratching between the \qty{20}{\micro\meter}  and \qty{70}{\micro\meter} scratch intervals, the normal force is observed to increase while the penetration depth, evidenced by the SEM micrograph, decreases (Figure \ref{fig:experiment}b). Although no diffraction map was collected between these intervals, the behaviour is attributed to the accumulation of material pile-up at the surface, ahead of the indenter, which must be displaced for the scratch to progress. The displaced pile-up is visible in the residual SEM micrograph and the in-situ diffraction maps. 

\subsection{X-ray diffraction}

\subsubsection{In-situ evolution of diffraction peaks}

\begin{figure}[h]
    \centering
    \includegraphics[width = 1.00\linewidth]{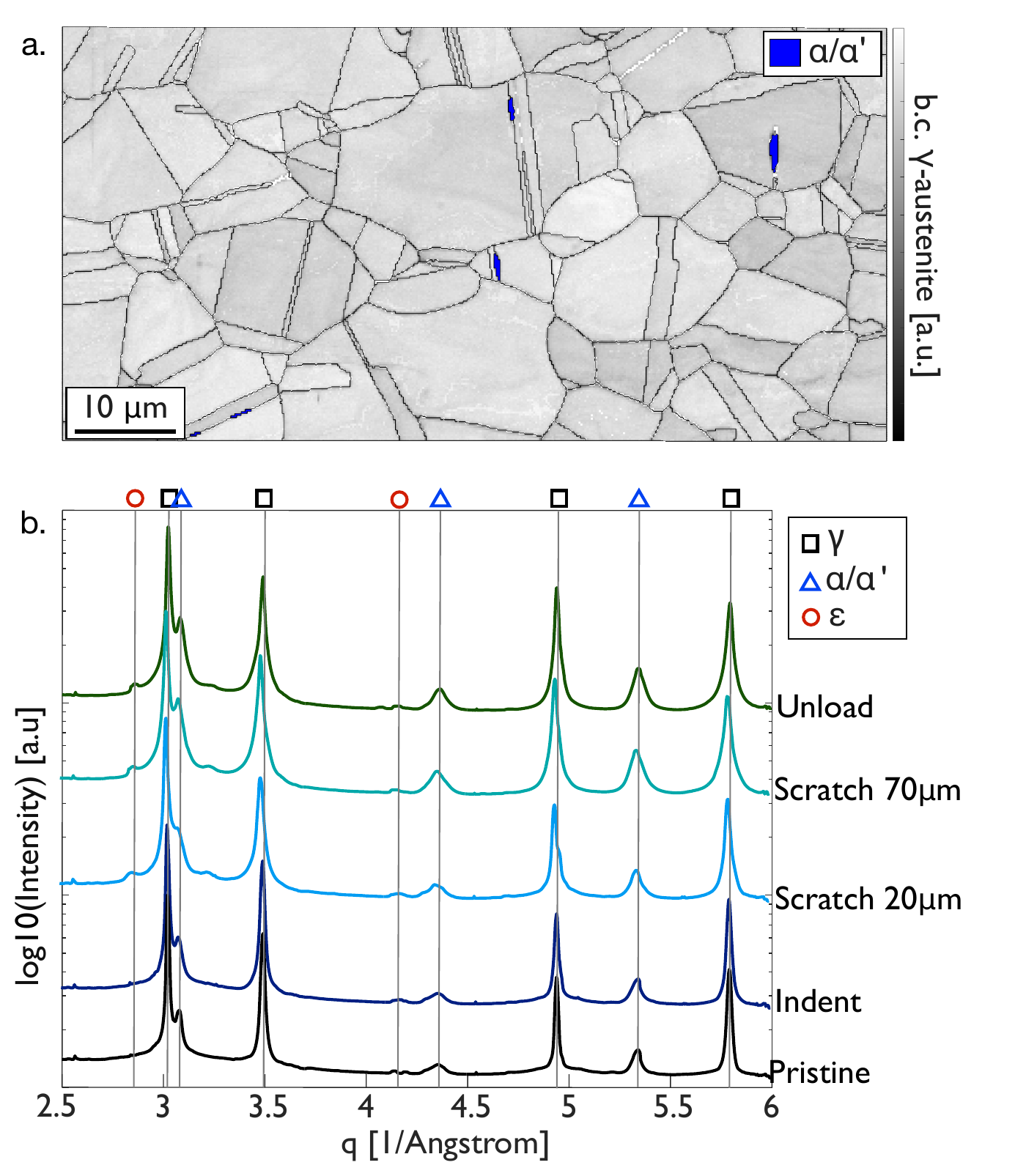}
    \caption{Phase identification for in-situ intervals. (a) EBSD phase map of the undeformed 316H stainless steel showing BCC/BCT $\alpha$-ferrite/$\alpha'$-martensite (blue) overlaid on the band contrast of the FCC $\gamma$-austenite matrix (greyscale). (b) Full azimuthally integrated diffraction intensity profiles obtained from a spatially averaged subset of deformed pixels for each test increment. Profiles are vertically offset for clarity. Peak positions for $\gamma$-austenite (squares), $\alpha$-ferrite/$\alpha'$-martensite (triangles), and $\varepsilon$-martensite (circles) are indicated.}
    \label{fig:peaks}
\end{figure}

Full azimuthal integration was performed on a spatially averaged subset of diffraction measurements within a deformed region. Integrated intensity profiles for each experimental increment are presented in Figure \ref{fig:peaks}. The measurement of the sample in the pristine condition confirms the presence of both the $\gamma$-austenite phase and an $\alpha$-ferrite/ $\alpha'$-martensite phase in the microstructure, consistent with EBSD measurements of the undeformed sample shown in Figure \ref{fig:peaks}a. Following indentation,  $\varepsilon$-martensite peaks appear, which increase in  intensity upon nanoscratching, indicating a strain induced $\gamma \ \rightarrow \varepsilon$ martensitic phase transformation. After scratching, the intensity of the $\alpha$-ferrite/ $\alpha'$-martensite peaks increases. Given the crystallographic similarity of BCT $\alpha'$-martensite and BCC $\alpha$-ferrite, these phases cannot be unambiguously distinguished by diffraction peaks alone; however, their increased intensity after deformation is interpreted as a strain induced $\gamma \rightarrow \alpha'$ martensitic transformation and herein referred to as $\alpha'$-martensite in the deformed intervals. The diffraction peaks show that both martensitic phases remain upon unloading.

\subsubsection{2D spatial mapping: Phase evolution}

\begin{figure*}[h]
    \centering
    \includegraphics[width = 0.9\linewidth]{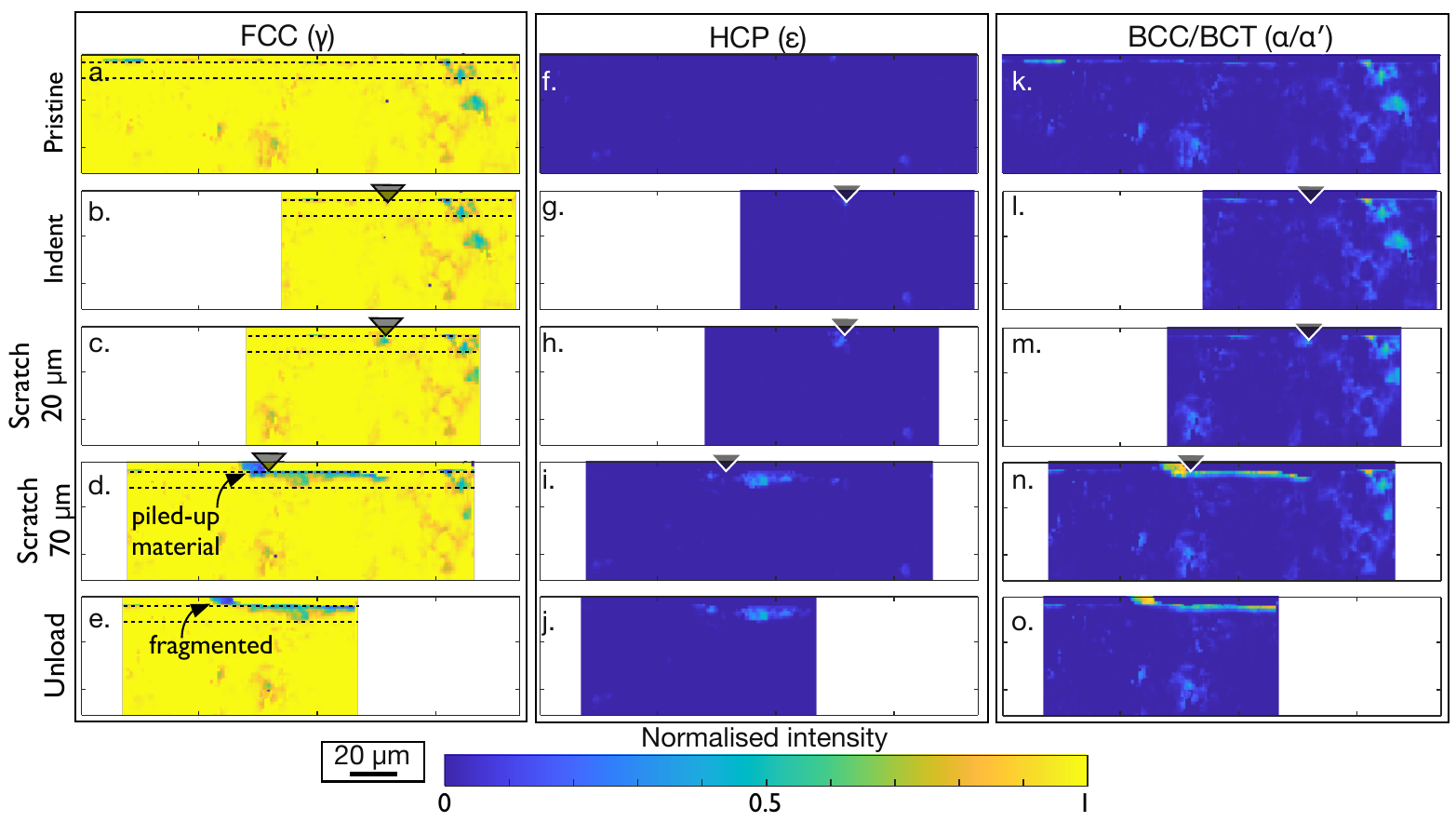}
    \caption{Spatial maps of the FCC $\gamma$-austenite (a - e), hcp $\varepsilon$-martensite (f - j), and the BCT $\alpha'$-martensite (k - o) phase fractions showing the progression of the $\gamma$ $\rightarrow$ $\varepsilon$ $\rightarrow$ $\alpha'$ transformation. Phase fractions were calculated from summed integrated peak areas and normalised to the total diffracted intensity from the three phases. For each phase, five maps (pristine, indent, scratch \qty{20}{\micro\meter}, scratch \qty{70}{\micro\meter} and unload) correspond to the force-displacement positions indicated by red circles in Figure \ref{fig:experiment}(b). The approximate position of the indenter tip relative to the sample is shown by the triangle and the dashed line indicates the subsurface deformed region. Propagated errors are provided in \ref{appendix 1}.}
    \label{fig:phase fractions}
\end{figure*}

Pixel-wise analysis of the diffraction patterns yielded 2D spatial maps of integrated intensity of each phase, $\gamma$-austenite bulk equivalent strain and dislocation density for the five experimental intervals. The relative intensity of each phase ($\gamma$-austenite, $\varepsilon$-martensite, and $\alpha$-ferrite/ $\alpha'$-martensite) was computed from the full azimuthally integrated diffraction intensity profiles, using the summed integrated intensity of two $\{hkl\}$ reflections per phase, normalised against the total summed intensity across all six reflections. Two-dimensional spatial distribution maps of the normalised phase intensity are shown in Figure \ref{fig:phase fractions}. In the pristine state, the maps show a microstructure dominated the $\gamma$-austenite phase, with a dispersed, heterogeneously distributed $\alpha / \alpha'$ phase distributed across the map, consistent with a pre-existing $\alpha$-ferrite phase. Following indentation, a localised region directly beneath the contact shows increase $\varepsilon$-martensite intensity, while the distribution of $\alpha / \alpha'$ remains largely unchanged, suggesting the deformation under indentation preferentially drives the $\gamma \rightarrow \varepsilon$ transformation. After the \qty{20}{\micro\meter} scratch increment, both $\varepsilon$ and $\alpha / \alpha'$  intensity increase in a localised region ahead of the indenter tip. The distribution of the increased $\alpha / \alpha'$  intensity within the heavily deformed region after scratching provides confidence that the deformation induced $\gamma \rightarrow \alpha'$ phase transformation has occurred. By the \qty{70}{\micro\meter} scratch increment, a fully established tribolayer is evident, characterised by a pronounced reduction in $\gamma$-austenite and a concurrent increase in both $\alpha'$ and $\varepsilon$ martensitic phases. The $\alpha'$-martensite is distributed across the entire scratch track and within the material pileup ahead of the indenter, while $\varepsilon$-martensite is concentrated within a localised subsurface region, at the later stages of the scratch track. Piled-up material is clearly visible ahead of the tip during the \qty{70}{\micro\meter} scratch interval, however, after unloading it appears to have detached from the surface, consistent with the SEM micrograph in Figure \ref{fig:experiment}.b). This displacement is likely a consequence of the lateral compliance in the indenter system, whereby the stored elastic strain energy in the loaded configuration is released upon unloading. This is akin to spring-back, leading to partial material detachment, and displacing the piled-up material.

\subsubsection{2D spatial mapping: Deformation evolution}

\begin{figure*}[h]
    \centering
    \includegraphics[width = 0.7\linewidth]{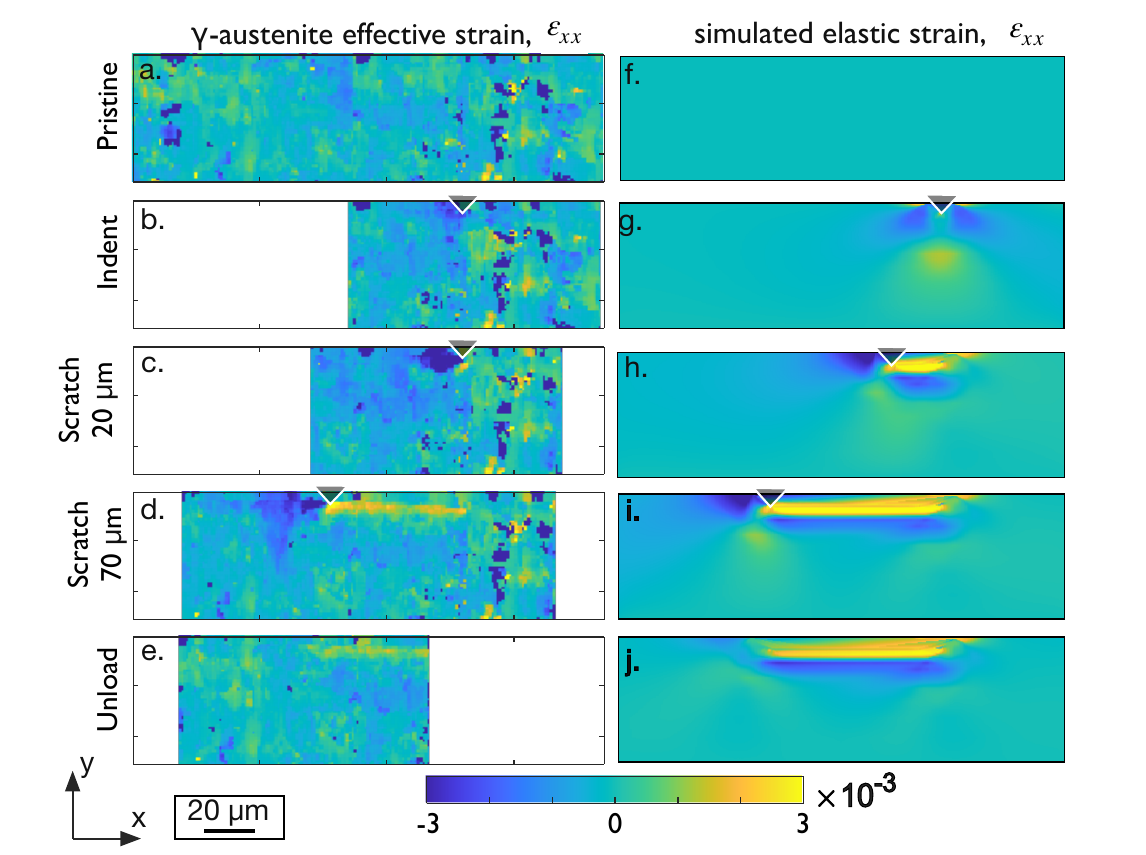}
    \caption{Spatial maps of effective $\gamma$-austenite strain $\varepsilon_{xx}$ (a - e), and the simulated elastic strain in the $x$-direction, $\varepsilon_{xx}$, predicted by the finite element model (f - j). The propagated errors for the experimentally measured $\varepsilon_{xx}$ are provided in Appendix \ref{appendix 1} Figure\ref{fig:errordeformation} (a - e). The triangle symbol represents the approximate position of the indenter tip relative to the sample.}
    \label{fig:strain}
\end{figure*}

\begin{figure*}[h]
    \centering
    \includegraphics[width=0.6\linewidth]{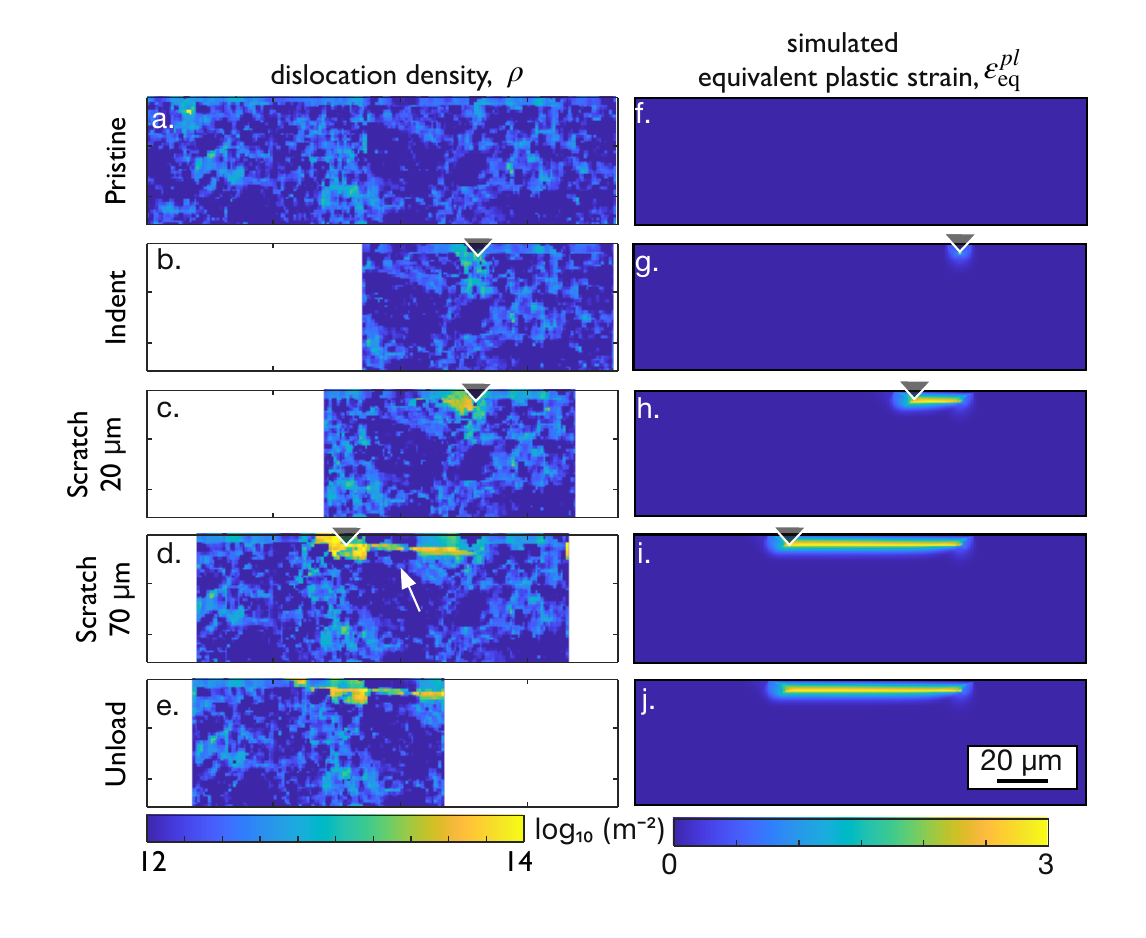}
    \caption{Spatial maps of the dislocation density, $\rho$, (a. - e.), estimated from diffraction peak broadening, presented as $log_{10}(\rho)$ in $m^{-2}$. f.-j. show the equivalent plastic strain (PEEQ) obtained from finite element model, representing the accumulated plastic deformation. Propagated errors for dislocation density are provided in Appendix \ref{appendix 1} Figure\ref{fig:errordeformation} (f-j). The triangle symbol represents the approximate position of the indenter tip relative to the sample.}
    \label{fig:dislocation}
\end{figure*}

Spatial distribution maps of the bulk strain component parallel to the scratch direction, $\varepsilon_{xx}$, and dislocation density, $\rho$, in the $\gamma$-austenite phase are presented in Figures \ref{fig:strain} and \ref{fig:dislocation}, respectively, along with finite element predictions for elastic strain and equivalent plastic strain. During indentation, $\varepsilon_{xx}$ exhibits a symmetric distribution of compressive and tensile strain about the contact axis (Figures \ref{fig:strain}.b), consistent with previous investigations \cite{kareer2025localised} and is reproduced here by the simulation in Figure \ref{fig:strain}.g. Figures \ref{fig:strain}.c and  \ref{fig:strain}.h show that upon scratching,  $\varepsilon_{xx}$ is compressive ahead of the contact and tensile in the wake of the scratch, reflecting the asymmetric loading conditions inherent to a sliding contact. This transition is most clearly resolved at the \qty{70}{\micro\meter} increment in Figures \ref{fig:strain}.d and \ref{fig:strain}.i, where the strain field is fully developed. The simulated elastic strain fields are in good qualitative and quantitative agreement with the experimental maps at each interval, confirming that directionally resolved bulk equivalent strains accurately capture the elastic strain field. The dislocation density maps in Figures \ref{fig:dislocation}.f–j show that plastic deformation is confined to a narrow band, close to the surface, with negligible plasticity extending into the far field. 

While this is also consistent with the distribution of the simulated equivalent plastic strain, $\varepsilon^{pl}_{eq}$, shown in Figures \ref{fig:dislocation}.i–l, the isotropic hardening model is unable to capture the non uniform distribution of plastic deformation across the scratch track. In the \qty{70}{\micro\meter} scratch interval, a segment of the wake shows that the plastic deformation is confined to a narrower region relative to the remainder of the scratch (see wake segment highlighted by the arrow Figure \ref{fig:dislocation}.d). This segment of the wake was attributed to the extensive accumulation of pile-up, arising from variable contact mechanics, as discussed in section  \ref{nanoscratch experiment}. The displacement-controlled boundary conditions using in the model were chosen to replicate target experimental displacements, and did not consider compliance of the instrument or capture this variation. While this explains the discrepancy between the experimental results and the model output in this segment, the consistency between model prediction and the experiment, outside of this region suggests that use of a simple, isotropic hardening model was justified, despite omitting the kinematic hardening responsible for pile-up formation.

\subsubsection{Correlation between martensitic phase transformations and deformation}

\begin{figure*}[hbt!]
    \centering
    \includegraphics[width=0.9\linewidth]{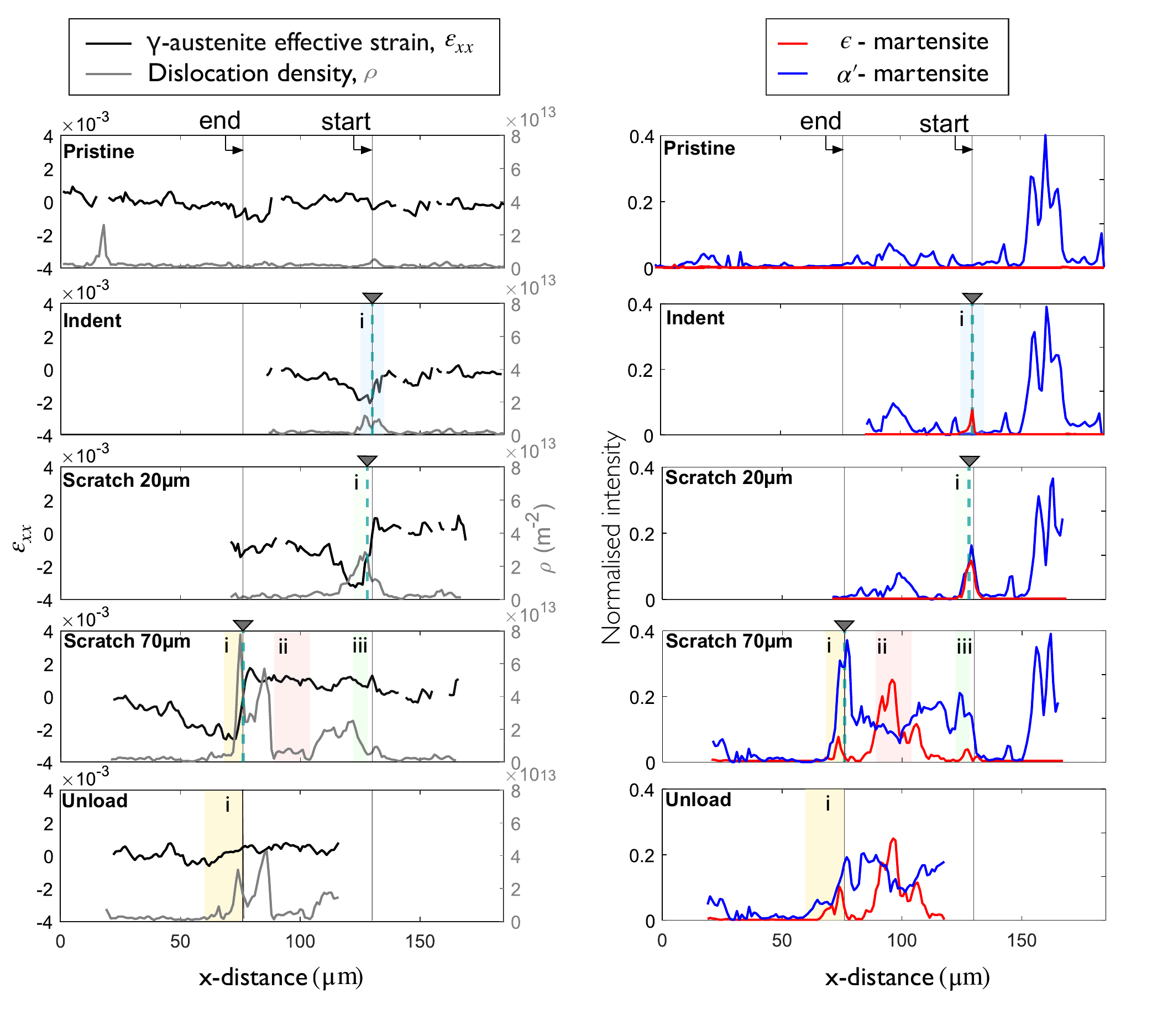}
    \caption{Line profiles extracted parallel to the scratch direction from spatial maps in Figures \ref{fig:phase fractions}, \ref{fig:strain} \& \ref{fig:dislocation}. Each profile averaged over an \qty{8}{\micro\meter} subsurface depth within the region bounded by the dashed lines in Figures \ref{fig:phase fractions}(a–e). Left panels show the effective $\gamma$-austenite strain, $\varepsilon_{xx}$ (black, left axis), and the total dislocation density, $\rho$ (grey, right axis). Right panels show the normalised integrated intensity of $\varepsilon$-martensite (red) and $\alpha'$-martensite (blue), plotted as a function of $x$-distance across the diffraction maps. Vertical dashed lines indicate the instantaneous contact position of the indenter; solid vertical lines mark the scratch start and end positions. Shaded regions denote spatially distinct zones discussed in the text: region i, the loaded contact region for the indent, \qty{20}{\micro\meter} and \qty{70}{\micro\meter} scratch intervals; region ii, a distinct wake segment with reduced plasticity; region iii, the unloaded wake region corresponding spatially to the loaded region i of the \qty{20}{\micro\meter} scratch interval.}
    \label{fig:corrlation x}
\end{figure*}

Figure \ref{fig:corrlation x} shows line profiles extracted parallel to the scratch direction. Each profile represents the average over an  \qty{8}{\micro\meter} subsurface depth, bound by the dashed horizontal lines shown in Figures \ref{fig:phase fractions}.a–\ref{fig:phase fractions}.e. Each profile is plotted as a function of the $x$-distance across the 2D spatial maps. For each loaded interval, the instantaneous contact position can be identified from the $\varepsilon_{xx}$ line profile; during indentation, the contact position corresponds to the peak compressive strain, $\varepsilon_{xx}$, while during scratching its position corresponds to the transition of the $\varepsilon_{xx}$ line profile from a compressive strain ahead of the contact, to tensile in the wake. The instantaneous contact positions are marked by the dashed vertical lines for the loaded intervals in Figure \ref{fig:corrlation x}; the scratch start and end positions are represented by solid vertical lines at $x$-distances \qty{130}{\micro\meter} and \qty{76}{\micro\meter}, respectively. The line profiles extracted from the pristine material provide baseline measurements for the pre-existing deformation and phase variation within the material. An elevated normalised $\alpha'$-martensite intensity is observed in the pristine material, at an $x$-distance between \qty{150}{\micro\meter} - \qty{170}{\micro\meter}; however, this lies outside the region of interest and has negligible influence on the subsequent analysis. 

Figure \ref{fig:corrlation x} is used to explicitly compare the effective strain and dislocation density in the $\gamma$-austenite phase with the normalised intensity of the martensitic phases, as the scratch contact evolves. Spatially distinct zones are highlighted in Figure \ref{fig:corrlation x}, corresponding to the instantaneous contact region under load during indentation and scratching, highlighted by the shaded regions i, while regions ii and iii represent the variable contact mechanics throughout the scratch. 

During indentation a compressive strain and elevated dislocation density are symmetrically distributed about the contact as highlighted by the blue shaded region i. The strain, $\varepsilon_{xx}$, reaches a maximum compressive value of $-2 \times 10^{-3}$  during indentation, while the dislocation density peaks symmetrically either side of the contact, reaching $\sim 1.2 \times10^{13}$\,m$^{2}$. A coincident, single peak is observed in $\varepsilon$-martensite normalised intensity, reaching a maximum of $\sim 0.07$, while no significant variation in $\alpha'$-martensite is observed relative to the pristine material.

During scratching, $\varepsilon_{xx}$ reflects the asymmetry of the sliding contact. The compressive strain peaks at $-3 \times 10^{-3}$, approximately \qty{5}{\micro\meter} ahead of the contact, for the \qty{20}{\micro\meter} scratch interval (left of the dashed line), while a tensile strain of $1 \times 10^{-3}$ develops behind the contact (right of the dashed line). The asymmetric distribution of $\varepsilon_{xx}$ is maintained for the \qty{70}{\micro\meter} scratch interval; the peak compressive strain ahead of the contact has a comparable magnitude to the \qty{20}{\micro\meter} interval, however, extends spatially over a larger distance ahead of the contact. In the wake of the contact, the tensile strain extends the full length of the established scratch track. The magnitude of the tensile strain, $\varepsilon_{xx}$, at the start of the scratch is consistent with the \qty{20}{\micro\meter} scratch interval and increases approximately linearly with scratch distance, reaching a peak tensile strain of $1.8 \times 10^{-3}$ immediately behind the contact. In the unloaded interval, the elastic strains are substantially reduced, with a compressive strain of $0.4 \times 10^{-4}$ retained ahead of the contact and a uniform tensile strain of $0.6 \times 10^{-3} $ retained within the wake.

The dislocation density retains a two-peak distribution either side of the contact for the \qty{20}{\micro\meter} scratch interval, but with marked asymmetry. In the wake of the contact, the dislocation density from the indentation is retained, while ahead of the contact it reaches a peak of $3 \times 10^{13}$\,m$^{-2}$, approximately three times that of indentation. An increase in the normalised intensity of both martensitic phases is observed ahead of the contact for the \qty{20}{\micro\meter} scratch interval, highlighted by the green shaded region i. Spatially, the peak intensity of both martensitic phases coincides with the peak dislocation density ahead of the contact, and extends ahead of the contact to a position that coincides with the peak compressive strain, $\varepsilon_{xx}$. 

Ahead of the contact of the \qty{70}{\micro\meter}, an elevated dislocation density coincides with a high normalised intensity of $\alpha'$-martensite as highlighted by the yellow shaded region i. Upon unloading, the dislocation density and the normalised intensity of the $\alpha'$-martensite appear to drop relative to the loaded contact position, (compare two yellow shaded regions i from the \qty{70}{\micro\meter} scratch interval and the unload interval). This is likely a result of the detached pile-up rather than recovery of the plastic deformation or a reduction in the fraction of the martensitic phase.  Within this region, the $\varepsilon$-martensite intensity peaks within the compressive strain field ahead of the contact, consistent with the behaviour observed for the \qty{20}{\micro\meter} scratch interval, and is retained upon unloading; the modified distribution of the  $\varepsilon$-martensite for the unloaded increment is similarly attributed to the displaced pile-up. 

The green shaded region iii of the \qty{70}{\micro\meter} interval corresponds spatially to the contact region (green shaded region i) of the \qty{20}{\micro\meter} interval, representing its unloaded state. Within this region, the compressive strain $\varepsilon_{xx}$ is relaxed while the dislocation density is retained. The $\alpha'$-martensite intensity is similarly retained, indicating that the $\gamma \rightarrow \alpha'$ transformation is irreversible once complete. In contrast to this behaviour, the $\varepsilon$-martensite intensity  at the start of the scratch has reduced to approximately one third of the value observed under the contact, approaching baseline levels further along the wake. 

A distinct segment of the wake, highlighted by the pink shaded region ii in the \qty{70}{\micro\meter} scratch interval, shows a reduction in $\alpha'$-martensite and an increase in $\varepsilon$-martensite that coincides with a reduction in the dislocation density. This region corresponds spatially to the segment discussed in sections \ref{nanoscratch experiment} and \ref{FE model method}  where an increase in the normal force is associated with the accumulation of piled-up material ahead of the indenter tip. Note that the effective strain consistently increases throughout the entire wake and is unaffected in this region, suggesting the distinction between these regions is attributed to the plastic deformation and pile-up formation. 

\begin{figure*}[h]
    \centering
    \includegraphics[width=0.5\linewidth]{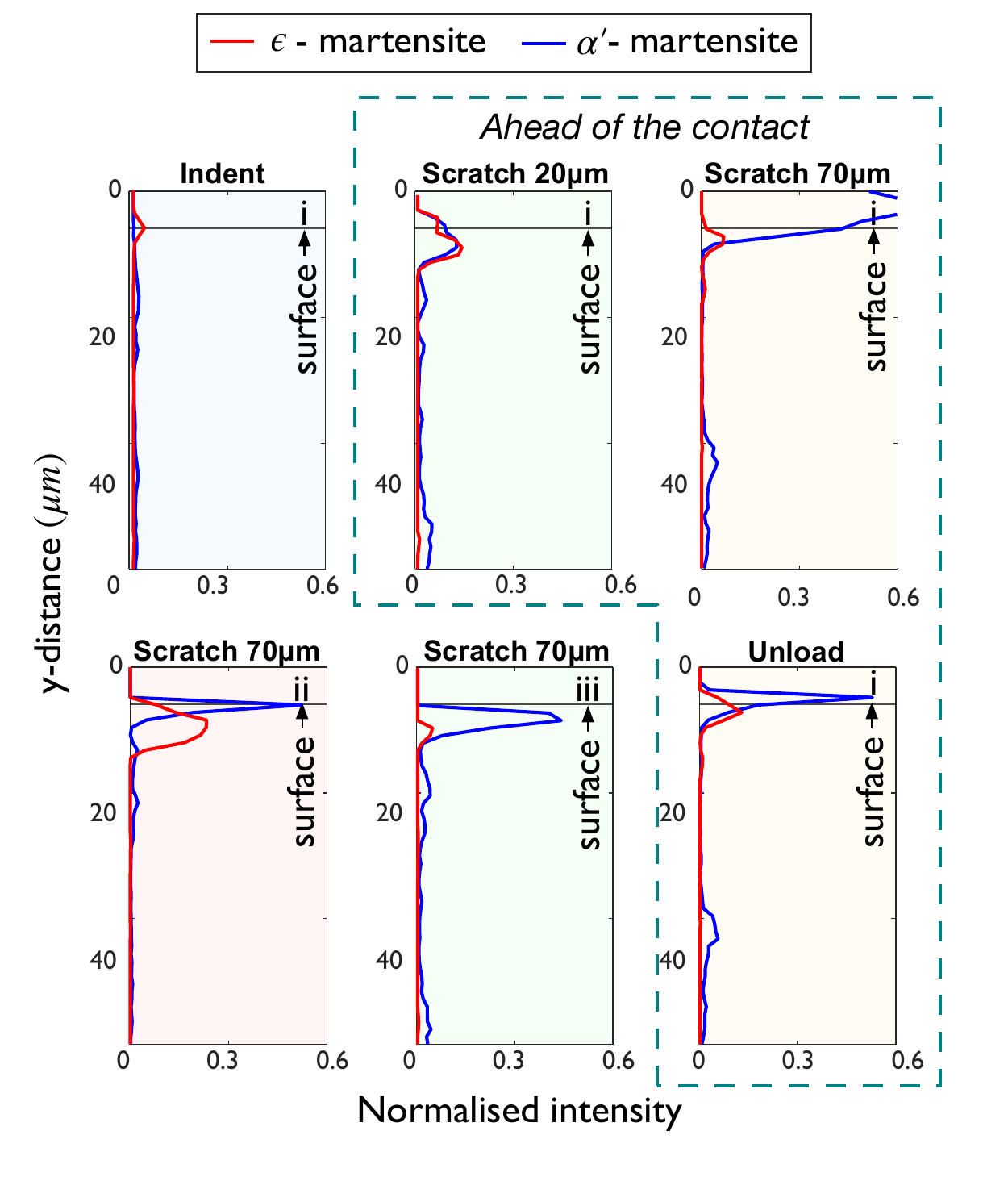}
    \caption{Vertical line profiles of normalised martensitic phase intensity as a function of y-distance below the sample surface, extracted from the spatial maps in Figure \ref{fig:phase fractions} and averaged over the shaded regions of interest identified in Figure \ref{fig:corrlation x}. Each profile is plotted with increasing y-distance representing increasing depth below the surface; the sample surface position is indicated by a horizontal dashed line. Profiles correspond to: the region ahead of the contact during indentation and scratching (top row), and the spatially distinct wake regions ii and iii during the \qty{70}{\micro\meter} scratch interval and upon unloading (bottom row). Red profiles represent $\varepsilon$-martensite; blue profiles represent $\alpha'$-martensite.}
    \label{fig:corrlation y}
\end{figure*}

The distribution of the martensitic phases above and below the surface is further examined through vertical line profiles. The line profiles in Figure \ref{fig:corrlation y} represent the shaded regions of interest highlighted in Figure \ref{fig:corrlation x}. Each profile is plotted as a function of $y$-distance which represents the vertical distance of the 2D spatial map from top to bottom. The sample surface is also indicated in Figure \ref{fig:corrlation y}. For the indent interval, vertical line profiles show that the $\varepsilon$-martensite is formed both above and below the surface, reflecting the symmetric pile-up of material around the contact.  Ahead of the contact during scratching, the distribution of martensitic phases within the pile-up is assessed. 

Ahead of the contact, at the initial stage of scratching, in the \qty{20}{\micro\meter} interval, the pile-up consists of both martensitic phases in approximately equal intensities; similarly, an equal intensity of both martensitic phases is observed subsurface; the subsurface intensity is larger than that in the pile-up for this initial scratch increment. For the \qty{70}{\micro\meter} interval and upon unloading, the $\alpha'$-martensite appears above the surface, with a considerably larger intensity, while the $\varepsilon$-martensite is at baseline values above the surface. This indicates extensive pile-up has accumulated above the surface, consisting of the $ \alpha'$-martensite, while subsurface, both martensitic phases are present. In contrast to the \qty{20}{\micro\meter} scratch interval, $ \alpha'$-martensite is the dominant phase.  This pile-up formation is further evidenced by the localised increase in the dislocation density directly ahead of the contact observed in Figure \ref{fig:corrlation x} (yellow shaded region i of the \qty{70}{\micro\meter} scratch interval), suggesting that the transformation in the pile-up coincides with substantial plastic deformation. Upon unloading, the intensity of the $\alpha'$-martensite is reduced above the surface which, as previously discussed, can be explained by the piled-up material being detached from the surface upon unloading.  

In both regions i and ii of the wake in the \qty{70}{\micro\meter} scratch interval, martensitic phases are only observed below the surface as the pile-up is displaced with the indenter. In all regions, the $\alpha'$-martensite is consistently concentrated in a band close to the surface, while $\varepsilon$-martensite is preferentially observed at a greater depth, below the surface. 
\section{Discussion}
\label{sec6}

The results show that the elastic strain field generated by a single sliding contact sufficiently activates both the intermediate $\gamma \rightarrow \varepsilon$ and $\gamma \rightarrow \alpha'$ strain induced martensitic transformations, and that they evolve sequentially as a direct result of the dynamic contact during sliding.  Unlike static indentation, where only $\varepsilon$-martensite is detectable in a localised region directly beneath the contact, sliding shows the presence of the $\alpha'$-martensite phase, indicating that this transformation requires the extensive shear strain inherent to sliding. 

Within the tribolayer of the established scratch track, two spatially distinct wake regions coexist; one region represents the unloaded wake of the scratch where the scratch was initiated (region iii) while the other region is formed within the established scratch wake, exhibiting a reduced dislocation density that corresponds with an increased normal force in the mechanical data (region ii). It is noted that the experiment was conducted as a series of quasi-static increments, with diffraction maps acquired while the stage was stationary rather than during continuous sliding. The diffraction maps therefore capture the accumulated deformation state at each interval rather than the evolving contact mechanics during active motion. The system also contained significant mechanical compliance, such that true steady-state scratching conditions were not maintained throughout the test. The reduced subsurface plasticity observed in region ii may therefore reflect a transient loading condition in which the applied normal force was transferred to the accumulated pile-up material rather than driving subsurface plastic deformation, though an additional measurement interval between \qty{20}{\micro\meter} and \qty{70}{\micro\meter} would be required for confirmation. Nonetheless, the evolution and spatial variation of deformation and dominant martensitic phase, captured  within a single experiment, provide an opportunity to examine the different transformation pathways. 

\begin{figure*}[hbt!]
    \centering
    \includegraphics[width=0.7\linewidth]{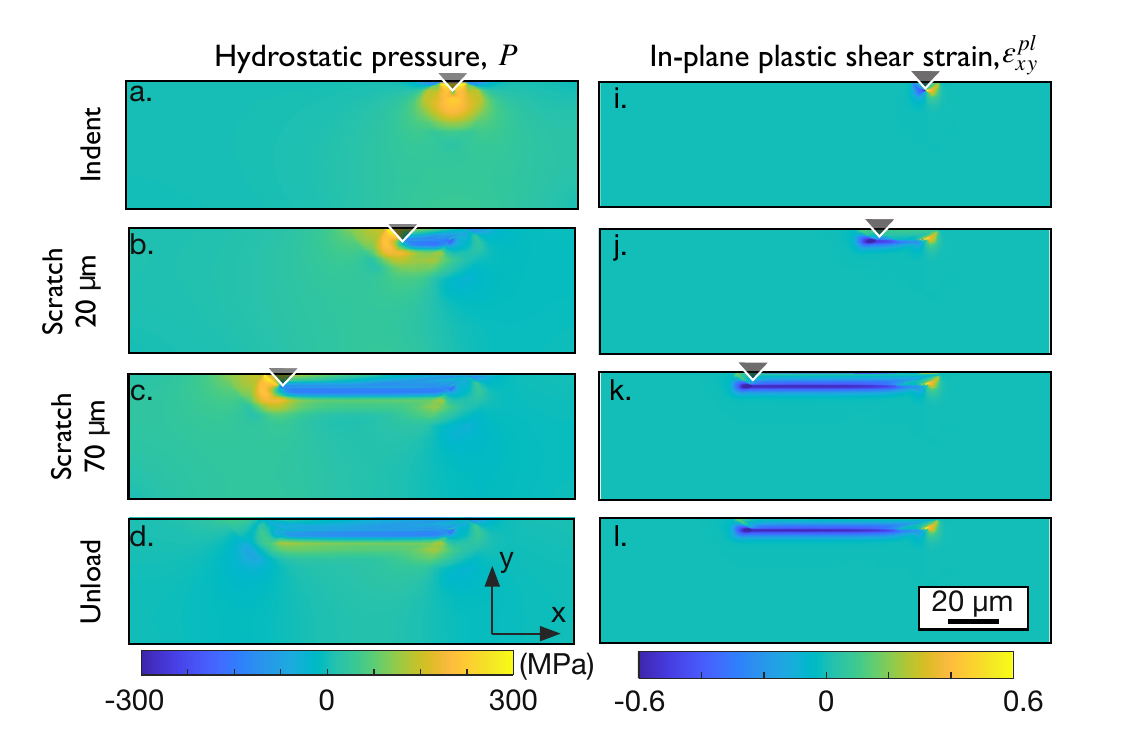}
    \caption{Finite element simulation predictions for the hydrostatic pressure, $P$, in MPa (a–d) and in-plane plastic shear strain component $\varepsilon^{pl}_{xy}$. The triangle symbol indicates the approximate position of the indenter tip relative to the sample surface. Note: The colour scale for hydrostatic pressure spans -300 to 300\,MPa, where a positive stress indicate hydrostatic compression.}
    \label{fig:pressure shear}
\end{figure*}

With the exception of the region exhibiting low dislocation density, the model predictions are valid throughout the contact and can be used to explain the observed transformation pathway from austenite to $\varepsilon$ martensite and subsequently to $\alpha'$ martensite. The in-plane plastic shear strain, parallel to the sliding direction, predicted by the finite element model is shown in Figure \ref{fig:pressure shear} i–l. While these shear strains provide the mechanical driving force for the strain induced martensitic transformations, the shear distribution alone does not explain their spatial distribution observed experimentally. In order to explain this, the volumetric distinction between the two transformations must be considered. 

Figure \ref{fig:pressure shear}.a-d show the simulated hydrostatic pressure. The regions of high hydrostatic pressure under the indentation and ahead of the contact, spatially coincide with regions where the $\varepsilon$-martensite is experimentally resolved (Figure \ref{fig:phase fractions}, Figure \ref{fig:corrlation x}). The volumetric dilatation from the $\gamma \rightarrow \alpha'$ is suppressed by the hydrostatic compression within these regions, hence the $\gamma \rightarrow \varepsilon$ is preferentially selected and evolves ahead of the contact as the scratch progresses. 

In the wake, the passage of the contact partially relieves this hydrostatic compression. Close to the surface, the hydrostatic field relaxes and the $\varepsilon$-martensite that formed ahead of the contact, completes the transformation to $\alpha'$-martensite through the sequential $\gamma \rightarrow \varepsilon \rightarrow \alpha'$ transformation, for which the $\varepsilon$-martensite provides nucleation sites. Deeper below the surface, the compression is only partially relieved, hence the dilatation of $\alpha'$-martensite remains penalised, and the $\varepsilon$-martensite is retained. This interpretation is supported by the depth distribution of the phases resolved in the wake; $\alpha'$-martensite is concentrated in a band close to the free surface where it is unconstrained, while $\varepsilon$-martensite  persists at a greater depth within the tribolayer (Figure \ref{fig:corrlation y}). This provides a mechanistic explanation for the spatial distribution of martensitic phases observed in austenitic stainless steel reported in \cite{emurlaev2022friction}. The spatial evolution of $\alpha'$-martensite appearing when $\varepsilon$-martensite is reduced, and similarly, the retention of $\varepsilon$-martensite  where $\alpha'$-martensite  does not develop, indicates that the $\alpha'$-martensite in the wake forms by transforming the pre-existing $\varepsilon$-martensite, being the result of a reverse transformation back to $\gamma$-austenite.

The piled-up material behaves differently. Although the finite element model does not accurately reproduce the pile-up, and the simulated pressure within it is therefore not relied upon here, the pile-up is by construction a free surface and does not sustain a confining hydrostatic stress. The dilatation penalty is consequently absent, and the high plastic shear strain accommodated in the piled-up material, evidenced by the elevated dislocation density ahead of the contact (Figure \ref{fig:dislocation}, Figure \ref{fig:corrlation x}), drives a direct $\gamma \rightarrow \alpha'$ transformation without an $\varepsilon$-martensite precursor. The $\alpha'$-martensite observed above the surface in the pile-up (Figure \ref{fig:phase fractions}, \ref{fig:corrlation y}) is therefore attributed to the direct transformation pathway. 

\begin{figure}[hbt!]
    \centering
    \includegraphics[width=1\linewidth]{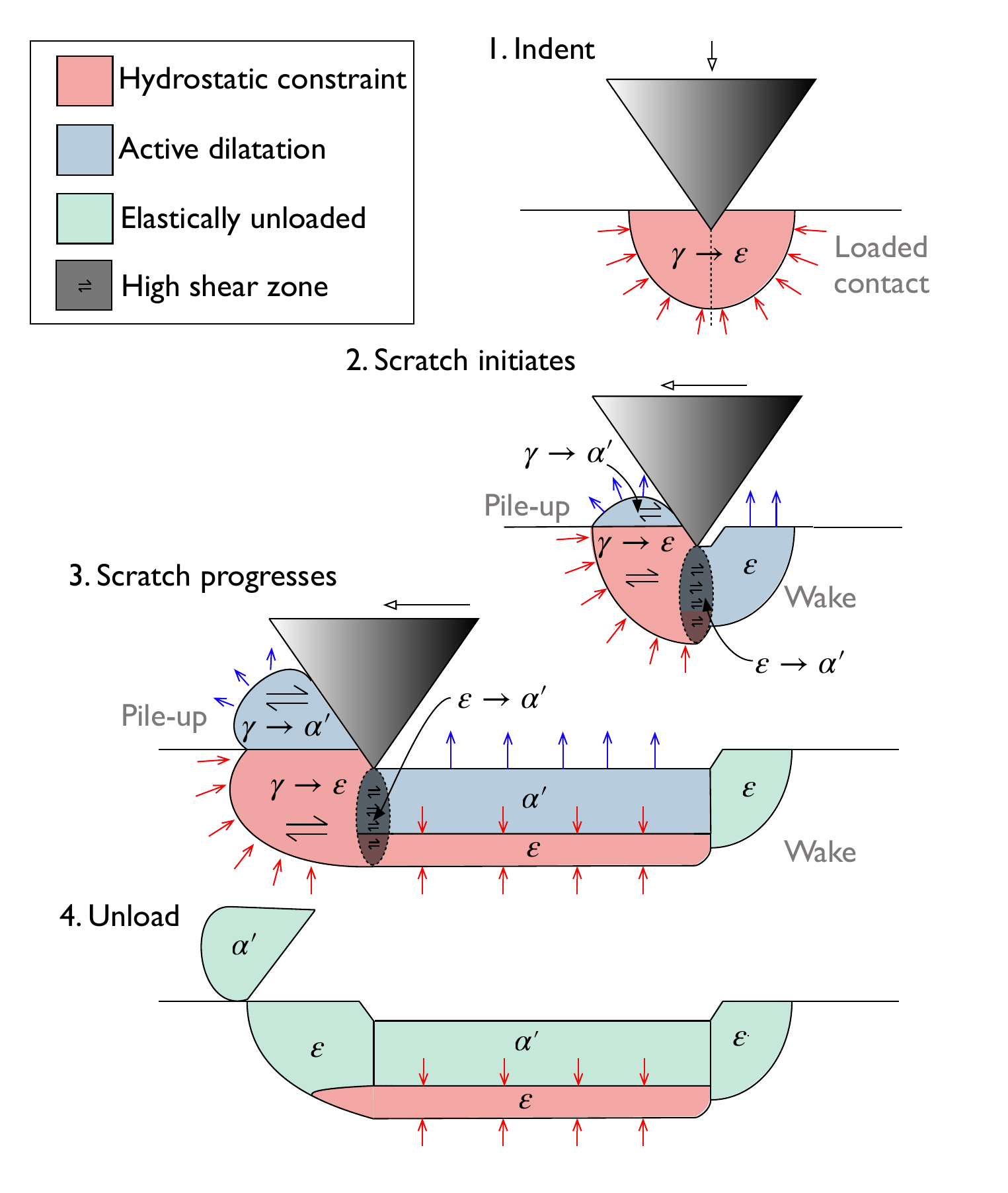}
    \caption{Proposed mechanistic relationship between imposed stress state, the associated deformation pathway and resultant martensitic phase for different spatial regions of the contact. Red regions represent the constrained material from hydrostatic compression; blue represents regions where the constraint is removed relative to the previous interval and material actively dilates upon $\alpha'$ formation; green represents elastically unloaded, static regions where the phase is retained from the previous increment. The high shear zone represents the dynamic zone beneath the frictional contact that subsequently transforms the $\varepsilon$-martensite ahead of the contact, thereby completing the sequential $\gamma \rightarrow \varepsilon \rightarrow \alpha'$ transformation.}
    \label{fig:mechanism}
\end{figure}

The transformation pathways are activated under distinct local stress conditions and evolve progressively as the contact advances, producing the spatially segregated phase distribution summarised in Figure \ref{fig:mechanism}. Four stages of the nanoscratch experiment are illustrated: (1) indentation, in which a compressive hydrostatic field beneath the contact promotes the volume-conserving $\gamma \rightarrow \varepsilon$ transformation within a constrained zone; (2) scratch initiation, in which piled-up material ahead of the contact undergoes a direct $\gamma \rightarrow \alpha'$ transformation driven by high plastic shear strain and the absence of hydrostatic constraint at the free surface of the pile-up; (3) steady-state scratch progression, in which the dynamic frictional shear zone completes the sequential $\gamma \rightarrow \varepsilon \rightarrow \alpha'$ pathway in the wake as the compressive constraint is relaxed by passage of the contact, while $\varepsilon$-martensite is retained at depth where compression is only partially relieved; and (4) unloading, in which $\alpha'$-martensite persists throughout the tribolayer and the intermediate $\varepsilon$-martensite is retained preferentially at greater subsurface depth, where the constraint has not been removed. The $\varepsilon$-martensite is also retained in the two extreme regions of the scratch track, at the start and end of the scratch; the associated deformation fields in these regions are equivalent to those under static indentation and not subjected to the sliding deformation to complete the sequential transformation. 

To provide a physically consistent explanation for the spatial co-occurrence of $\alpha'$- martensite and an elevated dislocation density in the $\gamma$-austenite, alongside the comparatively low dislocation density coexisting with the $\varepsilon$-martensite in the 316H studied here, the volumetric dilatation associated with the $\gamma \rightarrow \alpha'$ transformation is considered. This dilatation imposes a strain on the surrounding austenitic matrix, which must plastically deform to accommodate the volume change. The mechanistic consequence is that the $\gamma \rightarrow \alpha'$ transformation in the tribolayer is necessarily accompanied by subsurface matrix plasticity, which, in the context of galling and adhesive wear, may promote junction growth and facilitate material transfer in a manner not observed for the $\gamma \rightarrow \varepsilon$ pathway. In cobalt-based Stellite alloys, the predominant strain induced phase transformation under sliding contact is the $\gamma \rightarrow \varepsilon$ transformation, which produces a hexagonal martensite tribolayer without the dilatational misfit strain \cite{zhao2018comparative}. The present observations of the two competing pathways in 316H provide a mechanistic basis for interpreting this distinction; as the $\gamma  \rightarrow \varepsilon$ transformation proceeds without the matrix plasticity that accompanies $\alpha'$, the relative galling resistance of cobalt-based alloys over iron-based austenitic steels may not simply arise from the  hardness change due to the martensite rich tribolayer, but a difference in the spatial characteristics of subsurface deformation. This interpretation provides a physically grounded rationale for the development of cobalt-free hardfacing alloys in which the $\gamma \rightarrow \varepsilon$ transformation pathway is promoted and stabilised. This work showcases the utility of in-situ methods to directly quantify wear-relevant nanoscratch testing; future endeavours should explore other hardfacing alloys to ascertain whether the mechanistic pathway distinction observed here can be generalised. 


\section{Conclusions}
\label{sec7}

This study provides the first spatially resolved, simultaneous in-situ measurement of the $\gamma \rightarrow \varepsilon$ and $\gamma \rightarrow \alpha'$ transformation pathways alongside the evolving stress field beneath a single asperity sliding contact, demonstrated here on 316H stainless steel. 
A custom-built indenter was used to perform a controlled nanoscratch, with transmission X-ray nanodiffraction and finite element modelling to rationalise the resulting transformation from an evolving stress-field behaviour. 
By removing the spatial and temporal averaging inherent to post-mortem, macroscale wear testing, the nanoscratch geometry isolates the local strain state at each stage of contact evolution and directly correlates it with the relative phase distribution at that moment. These distributions allow the transformation pathway to be determined at distinct regions within the tribolayer. The key findings are as follows.

\begin{enumerate}
    \item Under static indentation, metastable $\varepsilon$-martensite forms within a localised region beneath the contact, coincident with a compressive hydrostatic stress field.

    \item Under dynamic scratching, the sequential $\gamma \rightarrow \varepsilon \rightarrow \alpha'$ pathway occurs in the wake of the contact, while a direct $\gamma \rightarrow \alpha'$ pathway occurs within the unconstrained pile-up ahead of the contact, indicating that the transition from indentation to sliding activates spatially distinct transformation routes.

    \item The transformation pathway selection is rationalised by the volumetric character of each transformation: the dilatation of $\alpha'$ is suppressed under hydrostatic compression and instead proceeds where shear strain is high and confinement is absent (the pile-up), or once the compressive field relaxes in the wake, allowing pre-formed $\varepsilon$ to convert.

    \item The co-localisation of $\gamma \rightarrow \alpha'$ with elevated dislocation density in the surrounding austenite indicates that this pathway is accommodated by, and contributes to, plastic damage in the matrix beyond the hardened tribolayer.

    \item This association between $\gamma \rightarrow \alpha'$ and matrix plasticity provides a plausible mechanistic contribution to the inferior galling resistance reported for Fe-based hardfacings relative to cobalt-based alloys, in which $\gamma \rightarrow \varepsilon$ is the dominant sliding-induced transformation.

    \item The combined in-situ diffraction and finite element approach developed here enables candidate cobalt-free hardfacing alloys to be screened by their propensity to favour $\gamma \rightarrow \varepsilon$ over $\gamma \rightarrow \alpha'$, and extension of this framework to multi-asperity contact conditions and alternative alloy chemistries is a natural next step.
\end{enumerate}

\subsection{Declaration of Competing Interest}

The authors declare that they have no known competing financial interests or personal relationships that could have appeared to influence the work reported in this paper.

\subsection{CRediT authorship contribution statement}

A. Kareer: Conceptualisation, Methodology, Data curation, Formal analysis, Funding acquisition, Investigation, Visualization, Writing - original draft, Writing - review \& editing. R.W. Kerr: Data curation, D. Craven: Data curation, A. Davydok: Data curation, D.M. Collins: Data curation, Methodology, Writing - review \& editing. C.A. Krywka: Conceptualisation.

\subsection{Acknowledgements}

AK acknowledges support from the Engineering and Physical Sciences Research Council under Fellowship grant EP/R030537/1 and the MPLS Returning carers fund. We acknowledge DESY (Hamburg, Germany), a member of the Helmholtz Association HGF, for the provision of experimental facilities. Parts of this research were carried out at PETRA III. [Beamtime was allocated for proposal  I-20241130].

\appendix
\section{Uncertainty quantification}
\label{appendix 1}

Propagated standard errors for all fitted diffraction quantities are reported as one standard deviation $(1\sigma)$. Uncertainties in effective $\gamma$-austenite strain were obtained by propagation of peak position fitting errors using the delta method. Uncertainties in dislocation density were obtained by propagation of peak fitting errors using the delta method; the standard error of the corrected peak breadth was propagated through the weighted linear fit to obtain the uncertainty in the slope, which was subsequently propagated to the dislocation density.

\begin{figure*}
    \centering
    \includegraphics[width=1\linewidth]{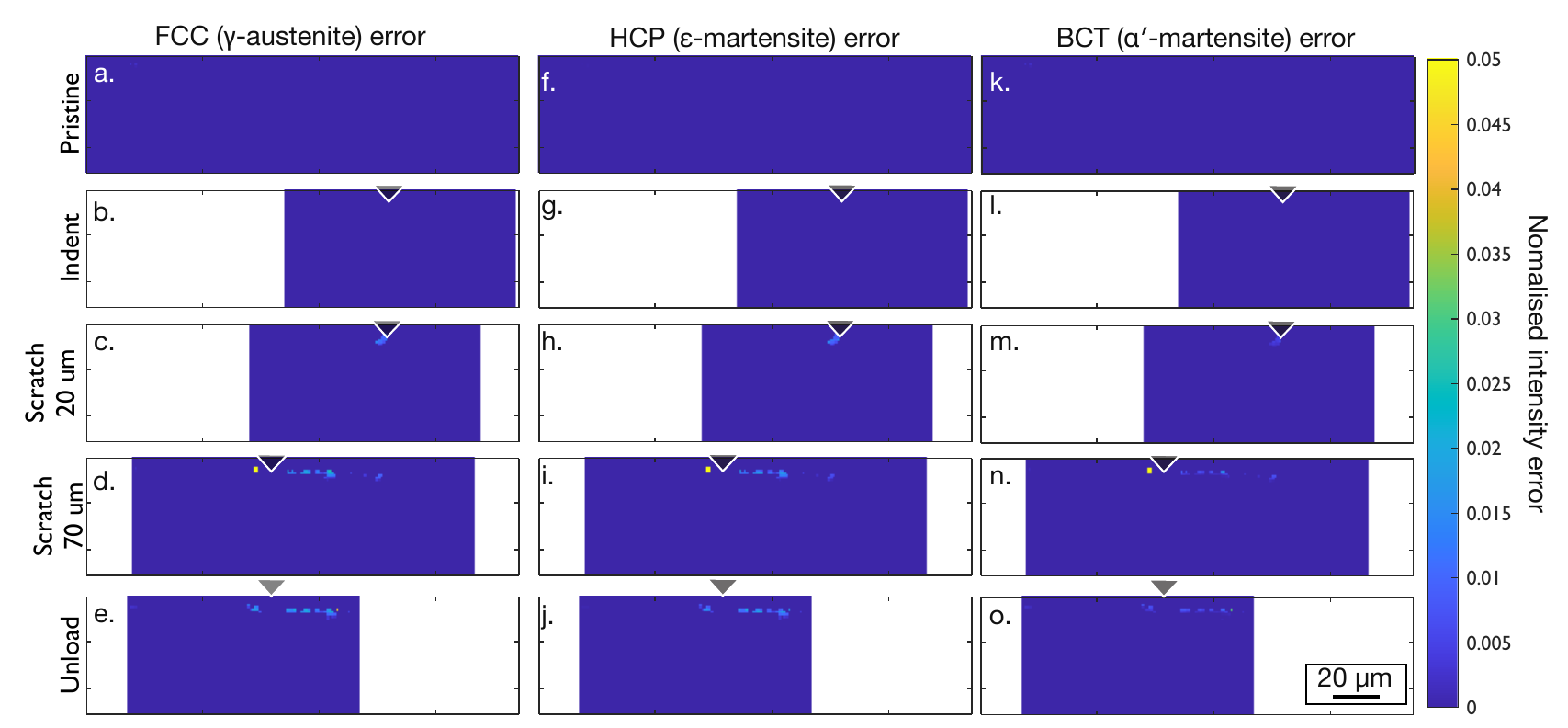}
    \caption{Propagated standard error for normalised integrated intensity, calculated using the delta method}
    \label{fig:errorintensity}
\end{figure*}

\begin{figure*}
    \centering
    \includegraphics[width=0.7\linewidth]{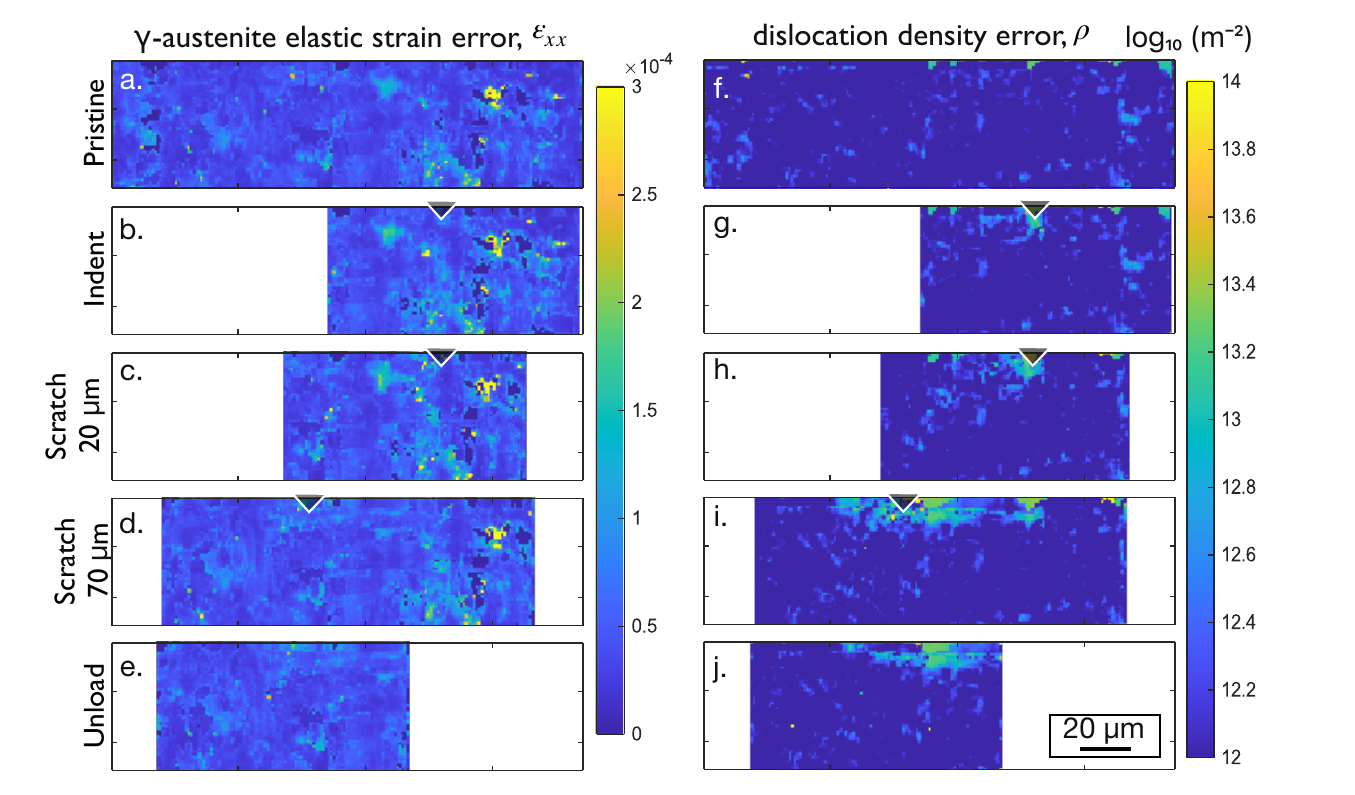}
    \caption{{Propagated standard error for lattice strain and dislocation density, calculated using the delta method}}
    \label{fig:errordeformation}
\end{figure*}

Spatial maps of the propagated standard errors for the normalised integrated phase intensity, the lattice strain components $\varepsilon_{xx}$, and the dislocation density, $\rho$, are presented in Figs. \ref{fig:errorintensity} and \ref{fig:errordeformation}, respectively.

\section{Finite element model}
\label{appendix 2}
The 316H stainless steel was modelled as an isotropic elastic-plastic solid with isotropic hardening. Elastic behaviour was defined by a Young's modulus of 193 GPa and a Poisson's ratio of 0.3, consistent with published values for austenitic stainless steel \cite{banerjee2023finite}. Plastic behaviour was defined by a tabulated true stress-plastic strain curve, provided in Table \ref{tab:plasticity}, obtained from \cite{do2021determination}. The curve was input as a piecewise linear hardening law in Abaqus. No rate-dependent or transformation-plasticity terms were included in the constitutive description; the model therefore represents the mechanical response of the parent austenitic phase and does not capture the volumetric or kinematic contributions of the martensitic transformations to the deformation field.

\begin{table}[hbt!]
\centering
\label{tab:plasticity}
\begin{tabular}{cc}
\toprule
plastic strain, $\varepsilon^{\mathrm{pl}}$ & Stress, $\sigma$ (MPa) \\
\midrule
0.000 & 290 \\
0.005 & 320 \\
0.010 & 340 \\
0.020 & 360 \\
0.050 & 390 \\
0.100 & 420 \\
0.150 & 440 \\
0.200 & 450 \\
0.250 & 460 \\[2pt]
0.300 & 469 \\[2pt]
0.400 & 482 \\[2pt]
0.500 & 493 \\
\bottomrule
\end{tabular}
\caption{Tabulated true stress-plastic strain data for 316H stainless steel used as input to the finite element model \cite{do2021determination}.}
\end{table}




\newpage

\bibliographystyle{elsarticle-num} 
\bibliography{References}

@article{Basham2015,
    author = {Basham, M. and Filik, J. and  Wharmby, M. T. and Chang, P. C. and El Kassaby, B. and Gerring, M. and Aishima, J. and Levik, K. and Pulford, B. C. and Sikharulidze, I. and Sneddon, D. and Webber, M. and Dhesi, S. S. and Maccherozzi, F. and Svensson, O. and Brockhauser, S. and  Náray, G. and Ashton, A. W. },
    title = {{Data Analysis WorkbeNch (DAWN)}},
    journal = {J. Synchrotron Rad.},
    year = {2015},
    pages = {853-858},
    volume = {22},
    doi = {10.1107/S1600577515002283}
}

@article{Filik2017,
    author = {Filik, J. and Ashton, A. W. and Chang, P. C. Y. and Chater, P. A. and Day, S. J. and Drakopoulos, M. and Gerring, M. W. and Hart, M. L. and Magdysyuk, O. V. and Michalik, S. and Smith, A. and Tang, C. C. and Terrill, N. J. and Wharmby, M. T. and Wilhelm, H.},
    title = {{Processing two-dimensional X-ray diffraction and small-angle scattering data in DAWN 2}},
    journal = {J. Appl. Cryst.},
    year = {2017},
    pages = {959–966.},
    volume = {50},
    doi = {10.1107/S1600576717004708}
}

@article{TODT2020425,
title = {Indentation response of a superlattice thin film revealed by in-situ scanning {X-ray} nanodiffraction},
journal = {Acta Mater.},
volume = {195},
pages = {425-432},
year = {2020},
doi = {10.1016/j.actamat.2020.05.056},
author = {J. Todt and C. Krywka and Z. L. Zhang and P. H. Mayrhofer and J. Keckes and M. Bartosik},
}

@article{kareer2016existence,
  title={The existence of a lateral size effect and the relationship between indentation and scratch hardness in copper},
  author={Kareer, Anna and Hou, XD and Jennett, Nigel M and Hainsworth, Sarah V},
  journal={Phil. Mag.},
  volume={96},
  pages={3396--3413},
  year={2016},
  doi = {10.1080/14786435.2016.1146828},
}

@article{rigney1997comments,
  title={Comments on the sliding wear of metals},
  author={Rigney, D. A.},
  journal={Tribol. Int.},
  volume={30},
  pages={361--367},
  year={1997},
  doi = {10.1016/S0301-679X(96)00065-5},
}

@article{Rigney1984,
  author    = {Rigney, D. A. and Naylor, M. G. S. and Rosenfield, A. R. and Jacobson, S.},
  title     = {Dislocation structures associated with sliding friction of copper},
  journal   = {Acta Metall.},
  year      = {1984},
  volume    = {32},
  pages     = {217--225},
  doi       = {10.1016/0001-6160(84)90083-3}
}

@article{emge2009effects,
  title={The effects of sliding velocity and sliding time on nanocrystalline tribolayer development and properties in copper},
  author={Emge, A. and Karthikeyan, S. and Rigney, D. A.},
  journal={Wear},
  volume={267},
  pages={562--567},
  year={2009},
  doi = {10.1016/j.wear.2008.12.102},
}

@article{yao2012correlation,
  title={Correlation between wear resistance and subsurface recrystallization structure in copper},
  author={Yao, B and Han, Z and Lu, K},
  journal={Wear},
  volume={294},
  pages={438--445},
  year={2012},
  doi = {10.1016/j.wear.2012.07.008},
}

@article{rau2021high,
  title={High diffusivity pathways govern massively enhanced oxidation during tribological sliding},
  author={Rau, Julia S and Balachandran, Shanoob and Schneider, Reinhard and Gumbsch, Peter and Gault, Baptiste and Greiner, Christian},
  journal={Acta Mater.},
  volume={221},
  pages={117353},
  year={2021},
  doi = {10.1016/j.actamat.2021.117353},
}

@article{hubner2003phase,
  title={Phase stability of {AISI} 304 stainless steel during sliding wear at extremely low temperatures},
  author={H{\"u}bner, Wolfgang and Pyzalla, A and A{\ss}mus, Kristin and Wild, E and Wroblewski, T},
  journal={Wear},
  volume={255},
  pages={476--480},
  year={2003},
  doi = {10.1016/S0043-1648(03)00164-9},
}

@article{greiner2016sequence,
  title={Sequence of stages in the microstructure evolution in copper under mild reciprocating tribological loading},
  author={Greiner, Christian and Liu, Zhilong and Strassberger, Luis and Gumbsch, Peter},
  journal={ACS applied materials \& interfaces},
  volume={8},
  pages={15809--15819},
  year={2016},
  doi = {10.1021/acsami.6b04035},
}

@article{greiner2018origin,
  title={The origin of surface microstructure evolution in sliding friction},
  author={Greiner, Christian and Liu, Zhilong and Schneider, Reinhard and Pastewka, Lars and Gumbsch, Peter},
  journal={Scripta Mater.},
  volume={153},
  pages={63--67},
  year={2018},
  doi = {10.1016/j.scriptamat.2018.04.048},
}

@article{brazil2021contribution,
  title={The contribution of plastic sink-in to the static friction of single asperity microscopic contacts},
  author={Brazil, Owen and Pethica, John B and Pharr, George M},
  journal={Proc. R. Soc. A:},
  volume={477},
  pages={20210502},
  year={2021},
  doi = {10.1098/rspa.2021.0502},
}

@article{pethica2023nanoindentation,
  title={Nanoindentation in more than one dimension--experimental challenges and opportunities},
  author={Pethica, John B},
  journal={Curr. Opin. Solid State Mater. Sci.},
  volume={27},
  pages={101100},
  year={2023},
  doi = {10.1016/j.cossms.2023.101100},
}

@article{kareer2020scratching,
  title={{Scratching the surface: Elastic rotations beneath nanoscratch and nanoindentation tests}},
  author={Kareer, Anna and Tarleton, Edmund and Hardie, Christopher and Hainsworth, Sarah V and Wilkinson, Angus J},
  journal={Acta Mater.},
  volume={200},
  pages={116--126},
  year={2020},
  doi = {10.1016/j.actamat.2020.08.051},
}

@article{kareer2025localised,
  title={Localised stress and strain distribution in sliding},
  author={Kareer, Anna and Demir, Eralp and Tarleton, Edmund and Hardie, Christopher},
  journal={Scr. Mater.},
  volume={263},
  pages={116662},
  year={2025},
  doi = {10.1016/j.scriptamat.2025.116662},
}

@article{jacobs2019insights,
  title={Insights into tribology from in situ nanoscale experiments},
  author={Jacobs, Tevis D. B. and Greiner, Christian and Wahl, Kathryn J and Carpick, Robert W},
  journal={MRS Bulletin},
  volume={44},
  pages={478--486},
  year={2019},
  publisher={Cambridge University Press},
  doi = {10.1557/mrs.2019.122},
}

@article{zeilinger2016situ,
  title={{In-situ observation of cross-sectional microstructural changes and stress distributions in fracturing TiN thin film during nanoindentation}},
  author={Zeilinger, Angelika and Todt, Juraj and Krywka, Christina and M{\"u}ller, Martin and Ecker, Werner and Sartory, Bernhard and Meindlhumer, Michael and Stefenelli, Mario and Daniel, Rostislav and Mitterer, Christian and others},
  journal={Sci. Rep.},
  volume={6},
  pages={22670},
  year={2016},
  doi = {10.1038/srep22670},
}

@article{rogers2020interaction,
  title={{The interaction of galling and oxidation in 316L stainless steel}},
  author={Rogers, Samuel R and Bowden, David and Unnikrishnan, Rahul and Scenini, Fabio and Preuss, Michael and Stewart, David and Dini, Daniele and Dye, David},
  journal={Wear},
  volume={450-451},
  pages={203234},
  year={2020},
  doi = {10.1016/j.wear.2020.203234},
}

@article{zhao2018comparative,
  title={{A comparative assessment of iron and cobalt-based hard-facing alloy deformation using HR-EBSD and HR-DIC}},
  author={Zhao, Chong and Stewart, David and Jiang, Jun and Dunne, Fionn P. E.},
  journal={Acta Mater.},
  volume={159},
  pages={173--186},
  year={2018},
  doi = {10.1016/j.actamat.2018.08.021},
}

@article{antony1983wear,
  title={{Wear-resistant cobalt-base alloys}},
  author={Antony, Kenneth C},
  journal={JOM},
  volume={35},
  pages={52--60},
  year={1983},
  doi = {10.1007/BF03338205},
}

@article{inglis1992performance,
  title={Performance of wear-resistant iron base hardfacing alloys in valves operating under prototypical pressurized water reactor conditions},
  author={Inglis, I and Murphy, E. V. and Ocken, H},
  journal={Surf. Coat. Technol.},
  volume={53},
  pages={101--106},
  year={1992},
  doi = {10.1016/0257-8972(92)90110-V},
}

@article{bowden2019understanding,
  title={Understanding the microstructural evolution of silicide-strengthened hardfacing steels},
  author={Bowden, D and Stewart, D and Preuss, M},
  journal={Mater. Des.},
  volume={161},
  pages={1--13},
  year={2019},
  doi = {10.1016/j.matdes.2018.09.015},
}

@article{cachon1996tribological,
  title={{Tribological qualification of cobalt-free coatings for pressurized water reactor primary-circuit gate valve applications}},
  author={Cachon, L and Denape, J and Sudreau, F and Lelait, L},
  journal={Surf. Coat. Technol.},
  volume={85},
  pages={163--169},
  year={1996},
  doi = {10.1016/0257-8972(95)02672-X},
}

@article{perdahciouglu2012macroscopic,
  title={{A macroscopic model to simulate the mechanically induced martensitic transformation in metastable austenitic stainless steels}},
  author={Perdahc{\i}o{\u{g}}lu, E. S. and Geijselaers, Hubertus J. M.},
  journal={Acta Mater.},
  volume={60},
  pages={4409--4419},
  year={2012},
  doi = {10.1016/j.actamat.2012.04.042},
}

@article{emurlaev2022friction,
  title={{Friction-induced phase transformations and evolution of microstructure of austenitic stainless steel observed by operando synchrotron X-ray diffraction}},
  author={Emurlaev, K and Bataev, I and Ivanov, I and Lazurenko, D and Burov, V and Ruktuev, A and Ivanov, D and Rosenthal, M and Burghammer, M and Georgarakis, Konstantinos and others},
  journal={Acta Mater.},
  volume={234},
  pages={118033},
  year={2022},
  doi = {10.1016/j.actamat.2022.118033},
}

@article{olson1972mechanism,
  title={{A mechanism for the strain-induced nucleation of martensitic transformations}},
  author={Olson, G. B. and Cohen, Morris},
  journal={J. Less-Common Met.},
  volume={28},
  pages={107--118},
  year={1972},
  doi = {10.1016/0022-5088(72)90173-7},
}

@article{mangonon1970martensite,
  title={The martensite phases in 304 stainless steel},
  author={Mangonon, Pat L and Thomas, Gareth},
  journal={Metall. Tran.},
  volume={1},
  pages={1577--1586},
  year={1970},
  doi = {10.1007/BF02642003},
}

@article{greenwood1965deformation,
  title={The deformation of metals under small stresses during phase transformations},
  author={Greenwood, Geoffrey Wilson and Johnson, R. H.},
  journal={Proc. R. Soc. A},
  volume={283},
  pages={403--422},
  year={1965},
  doi = {10.1098/rspa.1965.0029},
}

@book{hutchings2017tribology,
  title={{Tribology: Friction and Wear of Engineering Materials}},
  author={Hutchings, Ian and Shipway, Philip},
  year={2017},
  publisher={Butterworth-heinemann}
}

@article{vikstroem1994galling,
  title={{Galling resistance of hardfacing alloys replacing Stellite}},
  author={Vikström, Jussi},
  journal={Wear},
  volume={179},
  pages={143--146},
  year={1994},
  doi = {10.1016/0043-1648(94)90232-1},
}

@article{vannerem88chemistry,
  title={Chemistry of operating civil nuclear reactors},
  author={Vannerem, M},
  journal={Office for Nuclear Regulation, Nuclear Safety Technical Assessment Guide NS-TAST-GD-088 Revision},
year={2019},
  volume={2}
}

@misc{astm_g40_15,
  author = {{ASTM International}},
  title  = {{Standard Terminology Relating to Wear and Erosion, ASTM G40-15}},
  year   = {2015},
  note   = {ASTM International, West Conshohocken, PA},
  doi    = {10.1520/G0040-15}
}

@incollection{ahmed2017friction,
  title={Friction and wear of cobalt-base alloys},
  author={Ahmed, Rehan and de Villiers-Lovelock, Heidi},
  booktitle={Friction, Lubrication, and Wear Technology},
  pages={487--501},
  year={2017},
  publisher={ASM International},
  doi = {10.31399/asm.hb.v18.a0006390},
}

@article{persson2003influence,
  title={{The influence of phase transformations and oxidation on the galling resistance and low friction behaviour of a laser processed Co-based alloy}},
  author={Persson, Daniel H. E. and Jacobson, Staffan and Hogmark, Sture},
  journal={Wear},
  volume={254},
  pages={1134--1140},
  year={2003},
  doi = {10.1016/S0043-1648(03)00325-9},
}

@incollection{PaulCrook1990,
    author    = {P. Crook},
    title     = {{Cobalt and Cobalt Alloys}},
    booktitle = {Properties and Selection: Nonferrous Alloys and
                 Special-Purpose Materials},
    series    = {ASM Handbook},
    volume    = {2},
    publisher = {ASM International},
    address   = {Materials Park, OH},
    year      = {1990},
    pages     = {446--454},
    doi       = {10.31399/asm.hb.v02.a0001073},
}

@article{cockeram1999development,
  title={{Development of wear-resistant coatings for cobalt--base alloys}},
  author={Cockeram, B. V.},
  journal={Surf. Coat. Technol.},
  volume={120},
  pages={509--518},
  year={1999},
  doi = {10.1016/S0257-8972(99)00492-2},
}

@article{bastola2024experimental,
  title={{Experimental and numerical investigations of sliding wear behaviour of an Fe-based alloy for PWR wear resistance applications}},
  author={Bastola, Ajit and McCarron, Ruby and Shipway, Philip and Stewart, David and Dini, Daniele},
  journal={Wear},
  volume={540},
  pages={205186},
  year={2024},
  doi = {10.1016/j.wear.2023.205186},
}

@article{persson2003antigalling,
  title={{Antigalling and low friction properties of a laser processed Co-based material}},
  author={Persson, Daniel H. E. and Jacobson, Staffan and Hogmark, Sture},
  journal={J. Laser Appl.},
  volume={15},
  pages={115--119},
  year={2003},
  doi = {10.2351/1.1514218},
}

@article{persson2003effect,
  title={{Effect of temperature on friction and galling of laser processed Norem 02 and Stellite 21}},
  author={Persson, Daniel H. E. and Jacobson, Staffan and Hogmark, Sture},
  journal={Wear},
  volume={255},
  pages={498--503},
  year={2003},
  doi = {10.1016/S0043-1648(03)00122-4},
}

@article{rogers2024mechanisms,
  title={Mechanisms of elevated temperature galling in hardfacings},
  author={Rogers, Samuel R and Stewart, David and Taplin, Paul and Dye, David},
  journal={Wear},
  volume={558},
  pages={205564},
  year={2024},
  doi = {10.1016/j.wear.2024.205564},
}

@article{do2021determination,
  title={{Determination of true stress-strain curve of type 304 and 316 stainless steels using a typical tensile test and finite element analysis}},
  author={Do Kweon, Hyeong and Kim, Jin Weon and Song, Ohseop and Oh, Dongho},
  journal={Nucl. Eng. Technol.},
  volume={53},
  pages={647--656},
  year={2021},
  doi = {10.1016/j.net.2020.07.014},
}

@article{carrington2024evolution,
  title={{The evolution of subsurface deformation and tribological degradation of a multiphase Fe-based hardfacing induced by sliding contact}},
  author={Carrington, Matthew John and Daure, Jaimie L and Utada, Satoshi and Ratia-Hanby, Vilma L and Shipway, PH and Stewart, DA and McCartney, David Graham},
  journal={Mater. Sci. Eng. A},
  volume={892},
  pages={146023},
  year={2024},
  doi = {10.1016/j.msea.2023.146023},
}

@article{kim2000temperature,
  title={{The temperature dependence of the wear resistance of iron-base NOREM 02 hardfacing alloy}},
  author={Kim, Jun-Ki and Kim, Seon-Jin},
  journal={Wear},
  volume={237},
  number={2},
  pages={217--222},
  year={2000},
  doi = {10.1016/S0043-1648(99)00326-9},
}

@article{frictioncoeffgangopadhyay1993friction,
  title={Friction and wear behavior of diamond films against steel and ceramics},
  author={Gangopadhyay, A. K. and Tamor, M. A.},
  journal={Wear},
  volume={169},
  pages={221--229},
  year={1993},
  doi = {10.1016/0043-1648(93)90302-3},
}

@article{banerjee2023finite,
  title={{Finite element modelling of transient behaviours and microstructural evolution during dissimilar rotary friction welding of 316 austenitic stainless steel to A516 ferritic steel}},
  author={Banerjee, Amborish and da Silva, Laurie and Rahimi, Salaheddin},
  journal={J. Adv. Join. Process..},
  volume={8},
  pages={100167},
  year={2023},
  doi = {10.1016/j.jajp.2023.100167},
}

@article{agius2020microstructure,
  title={Microstructure-informed, predictive crystal plasticity finite element model of fatigue-dwells},
  author={Agius, Dylan and Al Mamun, Abdullah and Simpson, Chris A and Truman, Christopher and Wang, Yiqiang and Mostafavi, Mahmoud and Knowles, David},
  journal={Comp. Mater. Sci.},
  volume={183},
  pages={109823},
  year={2020},
  doi = {10.1016/j.commatsci.2020.109823},
}

@article{ungar1999contrast,
  title={The contrast factors of dislocations in cubic crystals: the dislocation model of strain anisotropy in practice},
  author={Ung{\'a}r, T and Dragomir, I and R{\'e}v{\'e}sz, {\'A} and Borb{\'e}ly, A},
  journal={Appl. Crystallog.},
  volume={32},
  pages={992--1002},
  year={1999},
  doi = {10.1107/S0021889899009334},
}

@article{kroner1958berechnung,
  title={Berechnung der elastischen Konstanten des Vielkristalls aus den Konstanten des Einkristalls},
  author={Kr{\"o}ner, Ekkehart},
  journal={Z. Phys.},
  volume={151},
  pages={504--518},
  year={1958},
}

@article{daure2025significant,
  title={Significant improvement in elevated temperature galling resistance of austenitic iron-based hard-facings through heat treatment},
  author={Daure, J and Carrington, M. J. and Koti, Daniel and Shipway, P. H. and McCartney, D. G. and Stewart, D. A.},
  journal={Wear},
  pages={206487},
  year={2025},
doi = {10.1016/j.wear.2025.206487},
}

@article{Shen2019Carbon,
title={Carbon content-tuned martensite transformation in low-alloy TRIP steels},
author={Yongfeng Shen and X. X. Dong and X. Song and N. Jia},
journal={Sci. Rep.},
year={2019},
volume={9},
doi={10.1038/s41598-019-44105-6}
}

@article{Bhushan1996Contact,
title={{Contact mechanics of rough surfaces in rribology: Single asperity contact}},
author={B. Bhushan},
journal={Appl. Mech. Rev.},
year={1996},
volume={49},
pages={275-298},
doi={10.1115/1.3101928}
}

@article{Stoyanov2017Scaling,
title={{Scaling Effects on Materials Tribology: From Macro to Micro Scale}},
author={P. Stoyanov and R. Chromik},
journal={Materials},
year={2017},
volume={10},
doi={10.3390/ma10050550}
}

@article{Cappella2022Editorial:,
title={{Editorial: Tribology and atomic force microscopy – Towards single asperity contact}},
author={B. Cappella and D. Spaltmann and M. Gee},
journal={Front. Mech. Eng.},
year={2022},
volume={8},
doi={10.3389/fmech.2022.853934}
}



\end{document}